\documentclass[11pt]{article}
\usepackage[T1]{fontenc}
\usepackage[utf8]{inputenc}
\usepackage[english]{babel}
\usepackage{amsmath,amssymb,amsthm}
\usepackage{graphicx}
\usepackage{booktabs}
\usepackage{float}
\usepackage[margin=1in]{geometry}
\usepackage[colorlinks=true,linkcolor=blue,citecolor=blue,urlcolor=blue]{hyperref}
\usepackage{csquotes}
\usepackage{endnotes}
\let\footnote\endnote
\makeatletter
\newcommand{\nlabel}[1]{\immediate\write\@auxout{\string\global\string\@namedef{nr@#1}{\the\c@endnote}}}
\newcommand{\nr}[1]{\@ifundefined{nr@#1}{??}{\csname nr@#1\endcsname}}
\makeatother

\title{Why Cassini's Oval Could Not Replace Kepler's Ellipse?}
\newif\ifanon
\anonfalse
\ifanon
\author{}
\else
\author{Pascal Descamps\\[0.3em]
\normalsize Laboratoire Temps-Espace, Observatoire de Paris\\
\normalsize 77 avenue Denfert-Rochereau, F-75014 Paris, France}
\fi
\date{}

\usepackage{titlesec}
\titleformat{\paragraph}[block]{\normalfont\normalsize\bfseries}{}{0pt}{}
\titlespacing*{\paragraph}{0pt}{1.4ex plus .4ex}{0.4ex}
\begin{document}
\maketitle

\begin{abstract}
\noindent In the later seventeenth century, Kepler's ellipse was not yet accepted. Ward claimed that Boulliau's conic hypothesis amounted to a Ptolemaic equant at the ellipse's empty focus, and built his own model on it. Around 1675--1680, Giovanni Domenico Cassini proposed another curve, the ``oval'', a genus-1 quartic with no closed-form Kepler equation; yet its eccentric anomaly shares exactly the leading coefficient of Kepler's, and its empty focus is, for the Sun, 38 times closer to a true equant than the ellipse's. Dynamically, Binet's formula shows that the force producing the oval under the area law departs from the inverse square by an index $\mathcal{D}(e)\sim3/(4e^2)$, depending on the oval's parameter $e$: no single exponent serves the whole family. This absence of universality, not of an area law, distinguishes it from the ellipse, and arguably explains why Cassini, the Acad\'emie and Leibniz kept the question open after the \emph{Principia} (1687).
\end{abstract}


\section{Historical context: Kepler's ellipse and its seventeenth-century rivals}
\label{sec:intro}

It would be natural to assume that the publication of the \emph{Astronomia Nova} in 1609, and of the \emph{Epitome Astronomiae Copernicanae} in the years 1618--1621, settled the question of the shape and the law of motion of the planets. It did nothing of the sort: for nearly half a century, the reception of Kepler's ``laws'' - the expression is itself anachronistic - remained hesitant, contested, and, frankly, largely circumvented, including by astronomers who claimed to follow him.\footnote{For surveys of this reception, see W.\ Applebaum, ``Keplerian astronomy after Kepler: researches and problems'', \emph{History of science}, 34 (1996), 451-504; Curtis Wilson, ``Predictive astronomy in the century after Kepler'', in René Taton and Curtis Wilson (eds), \emph{Planetary astronomy from the Renaissance to the rise of astrophysics, part A: Tycho Brahe to Newton} (Cambridge: Cambridge University Press, 1989), chap.~10; Bruce Stephenson, \emph{Kepler's physical astronomy} (New York: Springer, 1987).}

\paragraph{A real obstacle, not merely a prejudice.} The difficulty was not only philosophical. Kepler himself had failed to find an exact geometrical method for passing from mean anomaly to true anomaly using the ellipse's own elements alone, and neither he nor anyone after him, for more than a century, would succeed - for a reason we now understand precisely: Kepler's equation is transcendental, and admits no exact analytical solution. Seth Ward, professor of geometry at Oxford and future Anglican bishop, used exactly this language in the preface to his own \emph{Astronomia geometrica} (1656): ``\emph{methodo non naturali, sed peregrina}'' - ``by a method not natural, but foreign.'' Read in context, the words describe the laborious, trial-and-error procedure - ``\emph{perfalsae positionis repetitam operationem}'', a repeated operation of false position - by which Kepler himself, in his \emph{Commentaries on the Motions of Mars}, had had to determine his aphelia and eccentricities from observation, for want of any exact geometrical route.\footnote{Seth Ward, \emph{Astronomia geometrica; ubi methodus proponitur qua primariorum planetarum astronomia, sive elliptica sive circularis, possit geometricé absolvi} (London, 1656), ``Praefatio'' (unpaginated, sig.\ tt3): ``Nempe qui ratione Ellipseos planetariae speciem, posset ex observatis Geometricé determinare plane nescivit; quinimo perfalsae positionis repetitam operationem, methodo non naturali, sed peregrina, inventionem Apheliorum atque Excentricitatis aggressus est'' [``[Kepler], who did not know at all how to determine geometrically, from observation, the specific form of the planetary ellipse, undertook instead the discovery of the aphelia and eccentricities by a repeated operation of false position, by a method not natural, but foreign.''] Translation by the present author, checked directly against the 1656 print (Internet Archive, shelfmark \texttt{b30325717}). The same passage is quoted, in French translation and applied more broadly to the difficulty of solving Kepler's equation of the centre as such, in John L.\ Heilbron, \emph{Astronomie et églises}, trans.\ M.\ Mashaal (Paris: Belin/Pour la science, 2003), chap.~3, p.~121, citing ``Ward, \emph{Astr.\ geom.} (1656), Praefatio, tt3.''}\nlabel{n2} It is this impossibility of an exact, direct method - not mere nostalgic attachment to Ptolemaic astronomy - that explains the seventeenth century's long proliferation of alternative models meant to bypass the direct calculation of the elliptical sector.

\paragraph{Boulliau: a late and reluctant convert.} As early as 1633, well before he came around to the ellipse at all, Ismaël Boulliau confided to Gassendi everything he thought wrong with Keplerian physics: ``I cannot sufficiently marvel at Kepler's meditations in physics [\ldots] the solar magnetic virtue by which he has the planets attracted by the Sun is in no way demonstrable [\ldots] astronomy and geometry must reject as pure chimeras these hypotheses and these physical equations of motion which exclude geometrical demonstration. One must hold to this axiom, that celestial motions are simple circular motions, or composed of circular ones, and this is why I reject and repudiate his ellipses, unless they are described by an epicycle in the manner of Copernicus or Tycho.''\footnote{Boulliau to Gassendi, 21 June 1633 (BnF, FF.\,13050, fol.~258v-259r), quoted and translated in Robert A.\ Hatch, \emph{The Collection Boulliau (BN, FF.\,13019-13059): an inventory}, Memoirs of the American Philosophical Society, 117 (Philadelphia, 1982), pp.~xxvii-xxviii.}\nlabel{n3} Boulliau added at once, proof that he already had full command of the purely geometrical side of the question: ``it is moreover easy to demonstrate that an ellipse is described by a circle whose diameter equals the difference of the ellipse's axes, the planet being supposed to complete a double revolution on this epicycle while the centre of the epicycle completes only a single one on the deferent''\footnote{Hatch, \emph{Collection Boulliau} (ref.~\nr{n3}), pp.~xxvii-xxviii.} - an epicyclic construction not unrelated, allowing for the difference in framework, to the reciprocating diametral motion by which Kepler himself, independently, arrives at properties of the ellipse. Boulliau would not convert to the ellipse as the planets' real trajectory until the second half of the 1630s, under the influence of Gassendi's observations of the 1631 transit of Mercury.

\paragraph{Boulliau, Ward, and the equant regained: a subtler story than it appears.} In his \emph{Astronomia Philolaica} (Paris, 1645) - ``perhaps the most important work on planetary systems between Kepler's own and the \emph{Principia}'', in Newton's own words\footnote{\emph{The correspondence of Isaac Newton}, vol.~4, p.~60, n.~13, cited in Hatch, \emph{Collection Boulliau} (ref.~\nr{n3}), p.~xxxv.} - Boulliau finally admits the ellipse, on the strength of ``general considerations'' about celestial motion, by way of a geometrical construction of his own, his ``conic hypothesis'' (Book~I, chaps.~xiv--xvi), while devoting an entire chapter (Book~I, chap.~xii, \emph{An Sol moveat planetas?}) to dismantling Kepler's solar physics as undemonstrable. It is Seth Ward who, in a short and since-celebrated tract (\emph{In Ismaelis Bullialdi Astronomiae Philolaicae fundamenta, inquisitio brevis}, 1654),\footnote{Seth Ward, \emph{Idea trigonometriae demonstratae; item praelectio de cometis, et inquisitio in Bullialdi Astronomiae Philolaicae fundamenta} (Oxford, 1654; the \emph{inquisitio} written in 1653).} claims that this conic hypothesis amounts, purely and simply, to relocating a Ptolemaic equant - the very device Kepler had denounced and abolished in the \emph{Astronomia nova}, a purely computational artifice placing uniform motion about a point where neither body nor force resided - to the ellipse's vacant focus, the one the Sun does not occupy. Ward was mistaken: Curtis Wilson later showed that Boulliau's conic hypothesis is equivalent neither to Kepler's own nor to a simple empty-focus equant, and that Boulliau's calculation procedure was itself partly incoherent.\footnote{Curtis Wilson, ``From Kepler's laws, so-called, to universal gravitation: empirical factors'', \emph{Archive for history of exact sciences}, 6 (1970), 89-170; see also Hatch, \emph{Collection Boulliau} (ref.~\nr{n3}), pp.~xlv-xlvi, n.~12.} Ward nonetheless built, two years later, his own \emph{Astronomia Geometrica} (London, 1656),\footnote{Ward, \emph{Astronomia geometrica} (ref.~\nr{n2}). The copy consulted directly by the author is that of the Internet Archive (shelfmark \texttt{b30325717}).} explicitly founded on this empty-focus-equant model - convinced, this time on his own account, that he was not merely simplifying the calculation but stating the physical truth of the motion.\footnote{Hatch, \emph{Collection Boulliau} (ref.~\nr{n3}), p.~xlv.} The discrepancy between Kepler's own true anomaly and the ``pseudo-true'' anomaly of Ward's theory is, at most, $e^2/16$ - for the Sun ($e=0.033$ in Heilbron's normalisation), barely $14''$ of arc, undetectable by the instruments of the day, but not zero.\footnote{Heilbron, \emph{Astronomie et églises} (ref.~\nr{n2}), chap.~3, pp.~121-2; our own independent derivation, in a differing eccentricity convention, confirms this figure exactly in \S\ref{sec:ellipse-vs-oval} below.} Boulliau replied the following year (\emph{Astronomia philolaica fundamenta clarius explicata}, 1657): he corrects a small residual error (of order $2'$) meanwhile detected in his tables for Mars, rejects Ward's identification outright, and calls Ward's ``new methods'' for finding aphelia and nodes downright impracticable.\footnote{Hatch, \emph{Collection Boulliau} (ref.~\nr{n3}), p.~xlvi, citing Ismaël Boulliau, \emph{Astronomia philolaica fundamenta}, ``Monitum'', pp.~55-6.} Tested \emph{a posteriori} against 28 of Tycho's observations of Mars, Boulliau's corrected hypothesis in fact fit the data better than Kepler's own - which explains the considerable, if short-lived, fortune the \emph{Astronomia Philolaica} enjoyed among the following generation of astronomers.\footnote{Hatch, \emph{Collection Boulliau} (ref.~\nr{n3}), p.~xlvi.} Kepler, ``a very lucid man,'' had in any case always known that no equant could exactly reproduce his own ellipse.\footnote{Heilbron, \emph{Astronomie et églises} (ref.~\nr{n2}), chap.~3, pp.~121-2.}

\paragraph{A double reservation about Keplerian astronomy.} This constant concern, shared by Boulliau and Ward alike, to reduce the ellipse and the area law to a Ptolemaic equant reflects a double underlying reticence. On one hand, Kepler is reproached for reasoning as a \emph{physicist} rather than a \emph{geometer}: an astronomy founded on solar forces and magnetic ``virtues'' was judged foreign to the geometrical purity expected of the science of the stars; Cassini would go so far as to wage a genuine crusade against what he called the \emph{ageometria} of his predecessors.\footnote{Heilbron, \emph{Astronomie et églises} (ref.~\nr{n2}), chap.~3, pp.~122-3.} Ward himself, Savilian professor of astronomy at Oxford, claims exactly this programme when defending his faculty's teaching against the Puritan John Webster's criticisms: astronomy is taught there, he writes, ``as [Copernicus] left it, or as improved by Kepler, Bullialdus [Boulliau], our own Professor [Ward himself], and other partisans of the elliptical way,'' Oxford's scholars having laboured, among other mathematical work, to ``perfect the elliptical hypothesis, and render it geometrical.''\footnote{Seth Ward, in John Wilkins and Seth Ward, \emph{Vindiciae academiarum} (Oxford, 1654), pp.~46 and 30, reproduced in facsimile in Allen G.\ Debus (ed.), \emph{Science and education in the seventeenth century: the Webster-Ward debate} (London: Macdonald; New York: American Elsevier, 1970), pp.~43 and 223.} On the other hand - and this is probably the more decisive, more concrete reservation - Kepler's astronomy is scarcely practicable: no one knew how to pass from the time elapsed to the position on the ellipse. The area of an elliptical sector is elementary ($E-\varepsilon\sin E$ in Kepler's variable); it is its \emph{inversion}, finding the position from the area, that has no closed form - what came to be called Kepler's problem. The problem would remain open for an extraordinarily long time: more than a century after the \emph{Astronomia Nova}, Philippe de la Hire was still attempting it (\emph{Remarques sur le mouvement des planètes\ldots}, 1710), as was Jacques Cassini (\emph{Mémoire pour déterminer la première équation des planètes suivant l'hypothèse de Kepler}, 1719) - proof, if any were needed, that the reception of Kepler's laws never went without saying, and remained, in essence, an unsolved problem until the dawn of the eighteenth century.

\paragraph{Cassini's oval: another curve, for another name.} Nothing, in any case, obliged the ellipse's partisans to settle for that one figure: perhaps some other curve would serve better - an ovoid, say, of the kind Kepler himself had tried and rejected. Giovanni Domenico Cassini, after settling in Paris, ``driven by the blind desire to attach his name to a discovery that would carry his reputation to the most distant posterity,''\footnote{I.\,B.\ Cohen, ``Leibniz on elliptical orbits: as seen in his correspondence with the Académie royale des sciences in 1700'', \emph{Journal of the history of medicine and allied sciences}, 17 (1962), 72-82, as cited in B.\ Morgado and V.\ Soares, ``Kepler's ellipse, Cassini's oval and the trajectory of planets'', \emph{European journal of physics}, 35 (2014), 025009.}\nlabel{n16} thus proposed, around 1675, to substitute for Kepler's ellipses an entirely new class of ovals - defined as the locus of points whose \emph{product} (rather than sum) of distances to two fixed points is constant.\footnote{Heilbron, \emph{Astronomie et églises} (ref.~\nr{n2}), chap.~3, pp.~118-19.} The scorn that Cassini's astronomy attracted must not be read as a verdict on the oval - Nicolas Mercator spoke with open disdain of Cassini's \emph{lucubratiunculae}, little late-night scribblings,\footnote{Reported by Heilbron, \emph{Astronomie et églises} (ref.~\nr{n2}), chap.~3, pp.~123-4. Mercator's piece of 1670 concludes that its hypothesis departs from the truth about as much as Ward's approximation does: N.\ Mercator, ``Some considerations of Mr Nic.\ Mercator, concerning the geometrick and direct method of Signior Cassini for finding the apogees, eccentricities, and anomalies of the planets'', \emph{Philosophical transactions}, 5 (1670), 1168-75. Contemporary reactions to the oval are discussed in \S\ref{sec:reception}.} and Delambre, a century on, would not hesitate to invoke a form of ``charlatanism''\footnote{Heilbron, \emph{Astronomie et églises} (ref.~\nr{n2}), p.~124, reporting Jean-Baptiste Delambre, \emph{Histoire de l'astronomie moderne}, vol.~1 (Paris, 1821).} - but Mercator's target was Cassini's 1669 method for finding apogees and eccentricities, which predates the oval (c.~1675--80), so that his disdain cannot be a reaction to it. The ovals themselves, known today as ``Cassini ovals,'' earned their author compliments from Newton himself and were, for a time, preferred to Kepler's own in learned Parisian circles. The sculptor who later engraved these ovals on the monument raised to Cassini at the Paris Observatory chose, whether or not he knew it, an emblem of real aptness: Cassini's geometrical audacity, his ambition for renown, his debt to Kepler, and his indifference to the opinion of his peers.

No direct contemporary refutation of the oval seems to have been published. It would take Newton's and Gregory's series methods, and above all the demonstration, in the \emph{Principia} (1687), that the area law follows necessarily from a central force - whatever its particular law of decrease - for Kepler's physics to cease looking like one convenient hypothesis among others, and to become a theorem.

\bigskip
\noindent We follow this rival along two lines, one kinematic and one dynamical. Kinematically, if a point sweeps equal areas in equal times along the oval as seen from a focus, what relation links the mean anomaly to position on the curve, and how well does the empty focus serve as an equant? The oval turns out to be numerically almost indistinguishable from the Kepler ellipse of the same apsidal distances\footnote{Morgado and Soares, ``Kepler's ellipse, Cassini's oval and the trajectory of planets'' (ref.~\nr{n16}), quantify the relative discrepancy between the two curves, at equal eccentricity, at $10^{-3}$ to $10^{-5}$ for most solar-system planets, and $2\,\%$ only for Mercury, the most eccentric.}, and to share the leading coefficient of Kepler's own equation. Dynamically, what central force would produce a Cassini-oval orbit under the area law, and how far does it depart from the inverse square? Here the gap is considerable and precisely measurable. The contrast between the two answers frames the historical question of why the oval remained a live option for a generation.

The plan is as follows. Section~\ref{sec:reception} gathers what the sources allow us to say about the oval and its first readers, and ends on two questions that a calculation can address. The oval and its equivalent Keplerian eccentricity are recalled in \S\ref{sec:oval-geometry}. The kinematics (eccentric anomaly, generalised Kepler equation, Jacques Cassini's 1740 calculation) is treated in \S\ref{sec:kinematics} and Appendices~\ref{sec:appendix-proof}--\ref{sec:appendix-jc}, the dynamics (Binet's formula, the required force and its deviation from the inverse square) in \S\ref{sec:dynamics}, and the empty focus as an equant, compared with Ward's, in \S\ref{sec:equant}. Section~\ref{sec:history} returns to the reception with these results in hand, and Section~\ref{sec:conclusion} concludes. Readers chiefly interested in the historical argument may read Sections~\ref{sec:intro}, \ref{sec:reception}, \ref{sec:history} and~\ref{sec:conclusion} directly.

\section{Cassini's oval and its first readers}
\label{sec:reception}

The sources allow us to say a good deal about how the oval was received, and surprisingly little about why it was proposed. This section collects what is known - Cassini's own words, the equant tradition from which the oval seems to spring, and the testimony of the Académie and of Leibniz - and ends by putting two questions that the evidence cannot settle but a calculation can. The sections that follow are more computational; some results they establish (the discrepancies of a few arcseconds, the force law, the index $\mathcal{D}(e)$) are announced here by forward reference, and Section~\ref{sec:history} returns to the reception with them in hand.

\paragraph{The primary sources, and how few there are.} Almost nothing survives of Cassini's own reasoning behind the oval; the whole of this article's primary-source basis rests on two short texts, forty-seven years apart. The first is Giovanni Domenico Cassini's 1693 \emph{Traité de l'origine et du progrès de l'astronomie}, which introduces the curve - without ever calling it an ``oval'' - as a general hypothesis for planetary motion:
\begin{quote}
\noindent ``Sur ce que l'on avoit cy-devant reconnu par plusieurs observations que la vitesse réelle des planettes augmente à proportion qu'elles approchent du soleil, \& qu'elle diminuë à mesure qu'elles s'en éloignent; l'on a inventé une ligne pour servir d'orbite aux planettes. Cette ligne est une maniere d'ellipse dans laquelle les rectangles faits par les lignes tirées de la planette à l'un \& à l'autre foyer sont toûjours égaux; au lieu que dans les ellipses ordinaires ce sont les sommes des deux distances des foyers qui sont toûjours égales entr'elles.''\footnote{Giovanni Domenico Cassini, \emph{De l'origine et du progrès de l'astronomie et de son usage dans la géographie et dans la navigation} (Paris, 1693), p.~36 (Gallica, shelfmark bpt6k1510911j). ``Since it had previously been recognised, through several observations, that the planets' true velocity increases as they approach the sun and decreases as they recede from it, a line was invented to serve as the planets' orbit. This line is a kind of ellipse in which the rectangles formed by the lines drawn from the planet to the one and the other focus are always equal, whereas in ordinary ellipses it is the sums of the two distances to the foci that are always equal.'' Translation by the present author.}
\end{quote}
The impersonal ``on a inventé'' leaves it undecided whether Cassini claims priority or merely reports a construction already current; two independent secondary sources place his own detailed study of the curve around 1680, some years after this passage's implied original proposal (dated to around 1675 by I.\,B.\ Cohen).\footnote{Cohen, ``Leibniz on elliptical orbits'' (ref.~\nr{n16}); MacTutor History of Mathematics Archive, ``Giovanni Domenico Cassini'', University of St Andrews, accessed July 2026; Robert Ferréol, ``Cassinian oval'', \emph{mathcurve.com}, accessed July 2026.}\nlabel{n22} The second source is Jacques Cassini's account, in his \emph{Éléments d'astronomie} (1740), of his father's curve and of its use for the motion of the Sun.\footnote{Jacques Cassini, \emph{Éléments d'astronomie} (Paris, 1740), book~II, chap.~vi, pp.~149-51: ``Depuis l'observation exacte de la grandeur apparente des diamètres du Soleil, mon père a trouvé une autre courbe différente de l'ellipse, qui sert à représenter fort exactement les mouvements vrais du Soleil, \& ses diverses distances à la Terre. Il suppose que la Terre étant placée à l'un des foyers de cette courbe, le Soleil la parcourt par son mouvement propre, de manière que tirant de son centre aux deux foyers de la courbe, deux lignes droites, le rectangle fait sur ces deux lignes, soit toujours égal au rectangle fait sur la plus grande \& la plus petite distance du Soleil à la Terre.'' (``From the exact observation of the apparent size of the Sun's diameters, my father found another curve, different from the ellipse, which serves to represent very exactly the true motions of the Sun and its various distances from the Earth. He supposes that, the Earth being placed at one of the foci of this curve, the Sun traverses it by its own motion in such a way that, drawing from its centre two straight lines to the two foci, the rectangle on these two lines is always equal to the rectangle on the greatest and the least distance of the Sun from the Earth.'' Translation by the present author.) The same pages work out a numerical example for the Sun in which the empty focus serves as an equant.} Between the two lies a gap of essentially undocumented reasoning that no amount of technical reconstruction can fill; what follows is accordingly an accounting of how the curve \emph{behaves}, not a reconstruction of why Cassini chose it.

\subsection{Cassini and the rejection of physical reasoning}
\label{sec:cassini-rejection}

Nothing in the sources suggests that Cassini himself ever discussed any time law - area, equant, or otherwise - attached to his oval. In the 1693 text, no area law, equant or other time law is attached to the oval; the only worked application that survives is his son's of 1740, in which the empty focus is used as an equant (reconstructed in \S\ref{sec:jc-example}). This silence is unlikely to be accidental: as recalled in \S\ref{sec:intro}, Giovanni Domenico Cassini waged, by his own account, a campaign against what he called the \emph{ageometria} of his predecessors - a principled rejection of any \emph{physical} explanation (forces, magnetic ``virtues,'' causes) in favour of a purely geometric, descriptive astronomy.\footnote{Heilbron, \emph{Astronomie et églises} (ref.~\nr{n2}), chap.~3, pp.~118-23.} The area law, in Kepler's hands, was not a mere calculating convention: it was the direct consequence of a hypothesised motive force emanating from the Sun - exactly the kind of ``physicalist'' reasoning Cassini rejected on principle. It is unsurprising, then, that he never asked whether such a law, or any dynamical law at all, applied to his own curve: the question simply did not arise within the geometrical framework he deliberately occupied. Rejecting Kepler's physics was not, for Cassini, rejecting an area law that happened to be \emph{false for the oval}; it was rejecting the area law \emph{as such}, ellipse and oval alike.

\subsection{Ward's equant and d'Alembert's conjecture: a possible origin of the oval}
\label{sec:ward-dalembert}

If Cassini left no time law for his oval, where might the curve itself have come from? The nearest technical precedent is Ward's empty-focus equant, and d'Alembert's conjecture links the two.

\paragraph{Ward's scheme, and the oval.} Ward, according to J.\,L.\ Russell, ``accepted the equant law and was apparently the first to use it in practice''\footnote{J.\,L.\ Russell, ``Kepler's laws of planetary motion: 1609-1666'', \emph{The British journal for the history of science}, 2 (1964), 1-24, p.~17; see also Victor E.\ Thoren, ``Kepler's second law in England'', \emph{The British journal for the history of science}, 7 (1974), 243-56.}. His dedication, addressed jointly to Neile, Hevelius, Gassendi, Boulliau, and Riccioli and dated 1 August 1655, acknowledges Boulliau's help in these words: ``motus aequalitatem ad Axem Coni [\ldots] ad umbilicum alterum Ellipseos referendo adjuvisti'' [``you helped me by referring the equality of motion \ldots\ to the other focus of the Ellipse'']\footnote{Ward, \emph{Astronomia geometrica} (ref.~\nr{n2}), dedicatory epistle (Internet Archive \texttt{b30325717}, ff.~12-22); translation by the author.}. The principle is stated at the very opening of Book~I: ``Hujus Ellipseos, cùm focus alter sit sol [\ldots], super alterum interim focum, ita temperatur planetae cujusque motus, ut temporibus aequalibus, aequales illic angulos absolvat'' [``Of this ellipse, one of whose foci is the Sun \ldots, the motion of each planet is regulated, with respect to the \emph{other} focus, such that in equal times it traverses equal angles there'']\footnote{Ward, \emph{Astronomia geometrica} (ref.~\nr{n2}), \emph{Astronomia solaris}, book~I, part~I, chap.~i, p.~1 (Internet Archive \texttt{b30325717}, f.~41); translation by the author.} The equality of motion is thus assigned to the \emph{other} focus, as an angle. In Book~II, chapter~ii of the \emph{Astronomia geometrica} (``De Calculo Motus Solaris, ex Ellipsi specie data, \& dato Angulo Anomaliae simplicis''), Ward solves, triangle by triangle, the ``Anomalia aequata'' (true anomaly) from the ``Anomalia simplex'' (mean anomaly) and the given eccentricity, using the empty focus as a vertex of the solved triangle - exactly the scheme Jacques Cassini would attribute to his father for the oval, eighty-four years later (see above). The discrepancy this hypothesis introduces for the ellipse is, to leading order, $+(\varepsilon^2/4)\sin2M$ (\S\ref{sec:ellipse-vs-oval}), of maximum amplitude $\varepsilon^2/4$ - the $e^2/16$ of \S\ref{sec:intro} in Heilbron's normalisation: $14''$ for the Sun, several arcminutes for Mercury. This is very likely the connection behind d'Alembert's conjecture (see below), that Cassini reasoned from the near-constancy of the rectangle of focal radii in a low-eccentricity ellipse - the very principle underlying Ward's approximation.

A conjecture reported by d'Alembert in the \emph{Encyclopédie} a century later proposes a mathematical origin for the curve: in a low-eccentricity ellipse, the rectangle of the two focal radii is nearly constant - the very principle behind Ward's approximation, see above - and Cassini may have reasoned that this rectangle should be made \emph{exactly} constant on a purpose-built curve.\footnote{Jean le Rond d'Alembert, art.\ ``Cassinoïde'', in \emph{Encyclopédie, ou dictionnaire raisonné des sciences, des arts et des métiers}, vol.~1 (Paris, 1751), p.~633. The passage is quoted in full by Ferréol (ref.~\nr{n22}), and the present author has checked that full quotation directly; it fully supports the conjecture summarised here: ``Or M.\ Cassini paraît avoir raisonné ainsi: puisque le rapport des secteurs élémentaires aux angles correspondants est comme ce rectangle, il sera constant dans une courbe où le rectangle serait constant. Il a en conséquence imaginé la cassinoïde.'' Tome~I of the original \emph{Encyclopédie} appeared in 1751; the 1784 date given by Ferréol refers to a later reprint, from which the passage is quoted.} The conjecture is at least quantitatively coherent. On an ellipse of eccentricity $\varepsilon$ and semi-major axis $a$, the focal radii are $a(1\mp\varepsilon\cos E)$, so their product $a^2(1-\varepsilon^2\cos^2E)$ varies over the orbit by the relative amount $\varepsilon^2$ exactly: $2.8\times10^{-4}$ for the Sun, $4.2\,\%$ for Mercury. The rectangle is thus ``nearly constant'' precisely at the order $\varepsilon^2$ at which Ward's equant departs from the true motion (amplitude $\varepsilon^2/4$, \S\ref{sec:ellipse-vs-oval}), and the oval of the same apsides, whose rectangle is the constant $A^2-1=e^2$, is the curve obtained by freezing that product at its value on the apsides of the ellipse. This fixes the oval's parameter: the rectangle $b^2$ corresponds, for focal half-distance $c$, to $e=b/c=\sqrt{1-\varepsilon^2}/\varepsilon$, that is $e\approx59.3$ for the Sun - the value used throughout \S\S\ref{sec:kinematics}--\ref{sec:equant} - and this is precisely the condition Jacques Cassini states in 1740, where the rectangle equals the product of the Sun's greatest and least distances. Nothing in the numbers contradicts d'Alembert; nothing in the sources confirms him. If correct, this would make Ward's method not merely a shared intellectual climate but a direct technical source for the oval's very definition - an attractive hypothesis, but, a century removed and unsigned by Cassini, a conjecture rather than an established fact.

\subsection{The status of the area law before the \emph{Principia}, and Bertrand's theorem}
\label{sec:arealaw-status}

A second reason for Cassini's silence is that the area law had not yet acquired its later standing.

It should also be stressed that the conceptual apparatus needed even to pose the question correctly barely existed when Cassini advanced his proposal, around 1675, studying it in more detail around 1680.\footnote{Cohen (ref.~\nr{n16}), for the 1675 date. Two independent secondary sources situate the study of the curve around 1680: MacTutor (ref.~\nr{n22}) (``In 1680 he studied the Cassinian curve [\ldots] as part of a study of the relative motions of the Earth and the sun''); Ferréol (ref.~\nr{n22}) (``courbe étudiée par Cassini en 1680'').} The theorem that an observed area law necessarily signs the action of some central force - whatever its law of decrease - appeared in the \emph{Principia} only in 1687, close to when Cassini proposed his oval on either dating, and it took decades more for this Newtonian result to be fully absorbed into astronomical practice. As for Bertrand's theorem (1873), which singles out the inverse-square and linear forces as the only central forces giving closed orbits for all initial conditions, it is the natural tool for asking whether a single power law could serve the whole oval family; Jean Sivardi\`ere, the first to apply it to this question, notes that the theorem ``could have easily been discovered by Newton''\footnote{Jean Sivardi\`ere, ``Kepler ellipse or Cassini oval?'', \emph{European journal of physics}, 15 (1994), 62-4, p.~64.}\nlabel{n30} - yet no trace of such an argument appears before 1994. Bertrand's theorem does not, however, exclude a force producing one given oval; it is the $e$-dependence of that force, quantified in \S\ref{sec:dynamics}, that matters.

In the 1670s--1690s, the area law did not yet carry the status of universal physical necessity it would progressively acquire after the \emph{Principia}. For Boulliau as for Ward, decades after the \emph{Astronomia Nova}, the area law remained one calculating hypothesis among others - on a par with the Ptolemaic equant, to which it was in fact routinely reduced, the discrepancy between the two amounting to at most $\varepsilon^2/4$ in anomaly (\S\ref{sec:ellipse-vs-oval}), i.e.\ barely $14''$ for the Sun, a refinement few ``geometers'' of the time judged decisive. Nothing suggests Cassini, trained in that same culture of equants and hypothetical orbits, escaped this ordering of priorities. It was neither Cassini nor even Mercator who supplied the conceptual tool needed to settle the matter, but the following generation - Newton, and two centuries later, Bertrand and Sivardière. Cassini's oval is thus less an error its author could have detected than a proposal made within a framework - pure geometry, without any theory of force - in which the very question of its dynamical realisability could not yet be posed.

\subsection{Leibniz: fifteen years of measured doubt}
\label{sec:leibniz}

An independent source lets the question be tested directly: Leibniz's correspondence with the Académie royale des sciences, recovered and published by I.\,B.\ Cohen in 1962\footnote{Cohen, ``Leibniz on elliptical orbits'' (ref.~\nr{n16}).}, and studied in depth more recently by Paolo Bussotti\footnote{Paolo Bussotti, \emph{The complex itinerary of Leibniz's planetary theory}, Science Networks: Historical Studies 52 (Cham: Springer, 2015).}\nlabel{n32}. Leibniz is not a traditionalist astronomer wedded to equants: he is one of the two founders of the differential calculus, and, from 1689, one of the very first to attempt a central-force derivation compatible with Kepler's ellipse (\emph{Tentamen de motuum coelestium causis}) - as qualified a contemporary witness as one could ask for.

On 3 September 1700, newly elected a foreign associate of the Académie, Leibniz wrote to Fontenelle, its permanent secretary, asking directly ``whether Messrs.\ Cassini and de la Hire, and other excellent astronomers you have, are still content with Kepler's ellipses, or whether they believe that other lines answer better to the observations''\footnote{Leibniz to Fontenelle, 3 September 1700, cited in Cohen, ``Leibniz on elliptical orbits'' (ref.~\nr{n16}), 76-7.}. Fontenelle's reply, dated 8 December 1700 - thirteen years after the \emph{Principia} - is unambiguous:
\begin{quote}
\noindent M.\ Cassini abandoned Kepler's ellipse long ago \ldots\ to substitute for it another, in which \ldots\ it is the rectangles of these lines that are always equal. M.\ de la Hire does not appear to have settled on any curve, and I confess to you that I rather suspect it will turn out, in the end, that the celestial bodies actually describe none that is regular or exact. This irregularity \ldots\ seems to me more fitting to physics, though rather inconvenient for astronomy.\footnote{Fontenelle to Leibniz, 8 December 1700, cited in Cohen, ``Leibniz on elliptical orbits'' (ref.~\nr{n16}), 79.}
\end{quote}
\noindent Thirteen years after Newton's geometric theorem establishing that the area law necessarily signs a central force, the Académie's permanent secretary could still, without any sense of extravagance, publicly entertain the idea that the planets follow \emph{no} regular curve at all - a measure of how open a question the ellipse itself, the precondition for any Keplerian area law, remained in 1700 at the summit of European astronomy.

The same letter carries a second piece of information directly relevant to \S\ref{sec:dynamics}: it reports that ``M.\ Varignon \ldots\ has recently given us a general method for finding the different central forces that impel a planet,'' crediting Newton and Leibniz as ``the first, and indeed the only \ldots\ to have carried geometry far enough to discover these gravitations of the planets toward the sun''\footnote{Fontenelle to Leibniz, 8 December 1700, cited in Cohen, ``Leibniz on elliptical orbits'' (ref.~\nr{n16}), 81, itself quoting Varignon, ``Des forces centrales, ou des pesanteurs nécessaires aux planètes pour leur faire décrire les orbes qu'on leur a supposés jusqu'icy'', \emph{Histoire de l'Académie royale des sciences}, year 1700 (mémoires), 218-19.}. In other words, the very programme of this article - computing, for a hypothetical orbit, the central force law that would produce it - was already, in principle, Varignon's programme in 1700, some twenty years after Cassini advanced his oval and while it was still, as Section~\ref{sec:history} will recall, a live subject of debate at the Académie, explicitly regarding ``the orbits \ldots\ hitherto supposed'' - a phrase that can hardly exclude the oval itself. What \S\ref{sec:dynamics} contribute, two and a half centuries later, is not the question, already posed informally by Varignon, but its exact, explicit, quantified resolution.

Leibniz's interest in Cassini's oval in fact predates this 1700 exchange by nearly a decade, as Bussotti shows. In a letter to Huygens of April 1692, Leibniz is sceptical but undecided: ``In the meantime I find Kepler's ellipses very much to my liking, since they agree so well with mechanics, and perhaps the aberrations come from the actions of the planets upon one another''\footnote{Leibniz to Huygens, April 1692, in G.\,W.\ Leibniz, \emph{Sämtliche Schriften und Briefe} (LSB), series~III, vol.~5, pp.~288-9, cited in Bussotti (ref.~\nr{n32}), p.~69.} - a cautious position that does not rule out the rival hypothesis. Five years later, in October 1697, writing to de la Hire himself, the tone has shifted: ``I imagine that work will have been done on the new lines that M.\ Cassini substitutes for Kepler's ellipses, and that he will give physical causes for them. The great learning that M.\ Cassini possesses in astronomy leaves me in no doubt that he had considerable reasons for establishing them''\footnote{Leibniz to de la Hire, October 1697, LSB, III, 7, p.~618, cited in Bussotti (ref.~\nr{n32}), p.~69.}. Nine years further on, in his own planetary treatise, the \emph{Illustratio} (1706), Leibniz maintains the same reserve: he would have preferred these questions ``all submitted more carefully to observation, especially since the eminent astronomer Giovanni Domenico Cassini has proposed ovals of a new kind, and the illustrious de la Hire \ldots\ has been satisfied with no hypothesis''\footnote{Leibniz, \emph{Illustratio tentaminis de motuum coelestium causis} (1706), in \emph{Gesammelte mathematische Schriften}, ed.\ C.\,I.\ Gerhardt, vol.~6, 254-5, cited and translated in Bussotti (ref.~\nr{n32}), p.~68.}. Nineteen years after the \emph{Principia}, one of the ablest minds of his generation to judge the question on its merits still treated the ellipticity of orbits - not the area law alone, but ellipticity itself, the very precondition of the whole Keplerian construction - as a matter for observation to settle, not a theorem already secured. This sustained interest across nearly fifteen years (1692--1706), from cautious doubt to deference to methodological reserve, shows that Cassini's oval was, for Leibniz, never a hypothesis to be dismissed out of hand: a doubt that persists, shifts register, but never quite resolves in the ellipse's favour.

One further, more technical point bears directly on the status of the area law in Leibniz's own system. His planetary theory does not \emph{derive} the area law from a central force in the Newtonian sense; it \emph{postulates} it, through the hypothesis of a ``harmonic circulation'' of the ether carrying the planets, of which the area law is, as Bussotti notes, ``essentially, a postulate, not a proven result''\footnote{Bussotti (ref.~\nr{n32}), p.~11: ``to admit a harmonic circulation means, essentially, to postulate, not to prove, the area law.''}. Huygens himself, corresponding with Leibniz, confesses he does not understand how this harmonic circulation can reproduce the same ellipses that Newton obtains from inverse-square gravity and centrifugal force alone\footnote{Huygens to Leibniz, LSB, III, 5, p.~337, cited in Bussotti (ref.~\nr{n32}), p.~11.}. That the area law - the cornerstone of all post-Newtonian celestial mechanics, and the silent premise of every calculation in \S\ref{sec:dynamics} - could remain, for one of the era's greatest mathematicians, a purpose-built postulate rather than a fully assimilated dynamical consequence confirms, by a route entirely independent of the equant tradition discussed above, that its status as a universal physical necessity was established only gradually, not all at once with the 1687 publication of the \emph{Principia}.

\paragraph{Two questions.} What the sources leave open can be stated precisely. \emph{Kinematically}: if one postulates, by pure formal analogy with Kepler, that a point sweeps equal areas in equal times along the oval as seen from a focus, how close does the resulting motion come to Kepler's, and could the oval's empty focus serve as an equant? \emph{Dynamically}: what central force would produce a Cassini-oval orbit under the area law, and how far does it depart from the inverse square? The next four sections answer these questions.

\section{Cassini's oval: definition and equivalent Keplerian eccentricity}
\label{sec:oval-geometry}

\paragraph{Definition.} A Cassini oval is the locus of points $P=(x,y)$ whose \emph{product} of distances to two fixed points $F_1=(a,0)$ and $F_2=(-a,0)$ - the foci - remains constant, equal to $b^2$:
\[
\sqrt{(x-a)^2+y^2}\,\sqrt{(x+a)^2+y^2} = b^2.
\]
The kinship with the ellipse, the locus of points whose \emph{sum} of distances to the foci is constant, is plain - product in place of sum. Squaring gives the quartic Cartesian equation
\[
(x^2+y^2+a^2)^2 = b^4 + 4a^2x^2,
\]
and, in polar coordinates centred on the midpoint of the foci ($x=r\cos\varphi$, $y=r\sin\varphi$),
\begin{gather*}
r^4 - 2a^2r^2\cos2\varphi + a^4-b^4=0 \quad\Longrightarrow\quad \\ r(\varphi)^2 = a^2\cos2\varphi + \sqrt{b^4-a^4\sin^22\varphi}.
\end{gather*}
As for the ellipse, the shape of the curve depends only on the ratio $e=b/a$; taking the half focal distance $a=1$ as the unit of length,
\begin{equation}
r(\varphi)^2 = \cos2\varphi + \sqrt{e^4-\sin^22\varphi}. \label{eq:centred-polar}
\end{equation}

\begin{figure}[H]
\centering
\includegraphics[width=\textwidth]{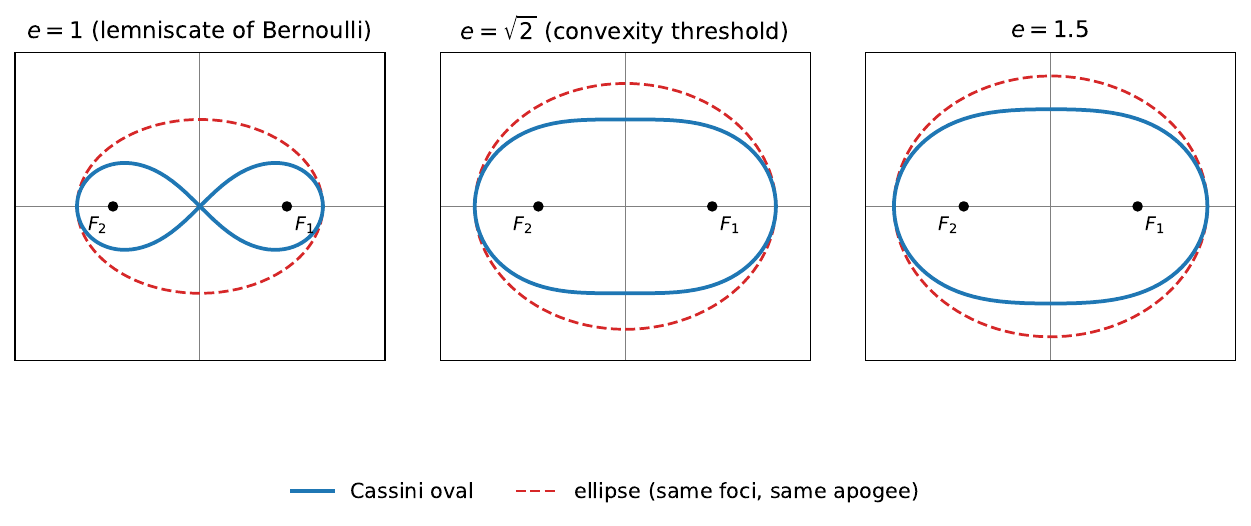}
\caption{Cassini ovals for $e=1$ (the lemniscate of Bernoulli), $e=\sqrt2$ (the threshold of convexity), and $e=1.5$ (solid line), compared with the ellipse of the same foci $F_1,F_2$ and the same apogee - that is, semi-major axis $A=\sqrt{1+e^2}$ and semi-minor axis $B=e$ (dashed line). At the lemniscate, the neck is zero and the curve self-intersects at the origin, far from any ellipse. At $e=\sqrt2$ the curve is only just convex, but still visibly ``squarer'' than the ellipse;}
\label{fig:cassini-panels}
\end{figure}

\paragraph{Shapes.} For $e<1$ the curve consists of two disjoint ovals, each surrounding one focus; $e=1$ gives the \textbf{lemniscate of Bernoulli}; for $1<e<\sqrt2$ the curve is a single, non-convex ``dumbbell''; for $e\geq\sqrt2$ it is convex - a genuine ``oval,'' increasingly close to a circle as $e$ grows (Figure~\ref{fig:cassini-panels}). Its extent along the focal axis is $2\sqrt{1+e^2}$ and the width of its neck is $2\sqrt{e^2-1}$.

\paragraph{Quantitative comparison with the Kepler ellipse.} To judge whether Cassini's oval can reasonably stand in for the Keplerian ellipse - Cassini's own apparent claim - a precise basis of comparison is needed: we take the ellipse of \emph{the same foci} $F_1,F_2$ (hence the same linear eccentricity, $a=1$) and \emph{the same apogee}, i.e.\ passing through the same point on the line of the foci as the oval. This ellipse has semi-major axis $A=\sqrt{1+e^2}$ and semi-minor axis $B=e$ (not to be confused with the oval's own ``neck,'' $2\sqrt{e^2-1}$, systematically narrower than the ellipse's diameter $2B$: this very narrowing is what makes the oval non-convex for $e<\sqrt2$).

\paragraph{The equivalent Keplerian eccentricity: an analytical formula.} Beyond curve-to-curve comparison, one may ask whether an \emph{analytical} equivalent of Keplerian eccentricity exists for Cassini's oval - a quantity intrinsic to the oval, independent of any particular choice of comparison ellipse. It does: a Cassini oval has, like an ellipse, two foci separated by $2a$ and a maximum extent along its major axis $A=\sqrt{a^2+b^2}$ (reached at $\varphi=0$). Defining eccentricity exactly as for the ellipse - the ratio of focal distance to semi-major axis, $\varepsilon=a/A$ - gives, with $a=1$ and $e=b/a$,\footnote{This definition, and the formula that follows, were proposed with equivalent notation by Sivardière, ``Kepler ellipse or Cassini oval?'' (ref.~\nr{n30}), 62-4 (his $c/\alpha$ is our $\varepsilon$, his ratio $d/c$ our $e$). Sivardière restricts his study to the curve's convex domain ($e>\sqrt2$), for which $\varepsilon$ does not exceed $1/\sqrt3\approx0.577$; the full domain of closed, non-self-intersecting ovals, $e\in[1,+\infty)$, including the non-convex ``dumbbell'' silhouettes, extends this result up to the true supremum $1/\sqrt2$, attained at the lemniscate ($e=1$).}
\begin{equation}
\varepsilon(e) = \frac{1}{\sqrt{1+e^2}}\, , \qquad e = \sqrt{\frac{1}{\varepsilon^2}-1}\ \text{(inverse relation).} \label{eq:eps-of-e}
\end{equation}
This function decreases strictly on the domain of closed, non-self-intersecting ovals, $e\in[1,+\infty)$, reaching its maximum precisely at the boundary $e=1$ (the lemniscate),
\[
\varepsilon_{\max} = \varepsilon(1) = \frac{1}{\sqrt2} \approx 0.7071,
\]
then decreasing toward $0$ as $e\to\infty$ (the circular limit). However elongated a non-self-intersecting Cassini oval may look, its equivalent Keplerian eccentricity can therefore never exceed $1/\sqrt2$ - a structural limit of the curve, distinct from the question of the actual resemblance of the \emph{shapes}.

\begin{figure}[H]
\centering
\includegraphics[width=0.72\textwidth]{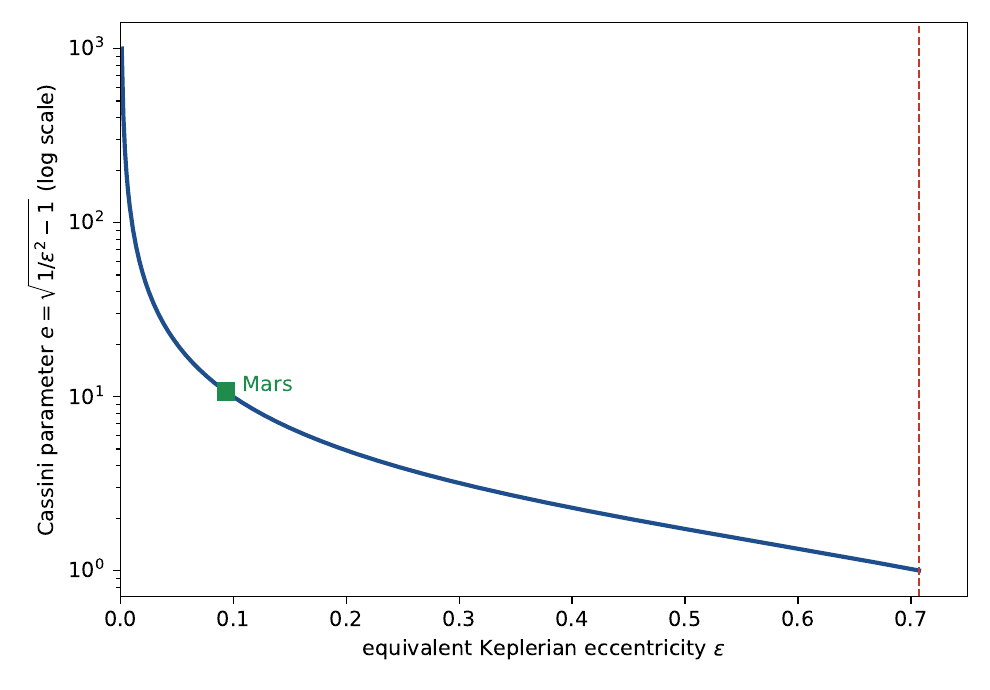}
\caption{Cassini parameter $e=\sqrt{1/\varepsilon^2-1}$ as a function of the equivalent Keplerian eccentricity $\varepsilon$ - the inverse of $\varepsilon(e)$, logarithmic scale for $e$. The curve diverges as $\varepsilon\to0$ (the circular limit, $e\to\infty$) and stops at $\varepsilon_{\max}=1/\sqrt2$, where $e=1$ (the lemniscate): beyond that value, no Cassini oval, whatever its parameter, can supply a larger equivalent eccentricity. The marked square shows, for reference, Mars's eccentricity ($\varepsilon\approx0.093$): matching it would require $e\approx10.7$ - confirming, by this inverse construction, that only ovals very close to circular can target realistic planetary eccentricities.}
\label{fig:cassini-e-vs-eps}
\end{figure}

Resemblance is real but slow to appear: the largest relative radial deviation between the oval and its comparison ellipse is still $25.5\,\%$ at $e=1.5$ ($\varepsilon\approx0.55$) and $2.0\,\%$ at $e=5$ ($\varepsilon\approx0.20$, comparable to Mercury). The relation $\varepsilon(e)$ can also be read in reverse: given a target equivalent Keplerian eccentricity $\varepsilon$, the Cassini parameter is $e=\sqrt{1/\varepsilon^2-1}$: $e\approx59.3$ for the Sun ($\varepsilon=0.01685$) and $e\approx4.76$ for Mercury ($\varepsilon\approx0.206$). Figure~\ref{fig:cassini-e-vs-eps} plots this inverse function.

\section{Kinematics: the oval's ``equation of the centre''}
\label{sec:kinematics}

\subsection{The oval seen from a focus}
\label{sec:quartic}

Everything that follows requires a focus-centred coordinate system, set up here.

Place the pole at the focus $F_2=(1,0)$ and let $\rho$ be the distance to that focus, $\varphi$ the corresponding polar angle. A point with coordinates $(1+\rho\cos\varphi,\rho\sin\varphi)$ is at distance
$$r_2=\rho, \qquad r_1^2=(2+\rho\cos\varphi)^2+\rho^2\sin^2\varphi = \rho^2+4\rho\cos\varphi+4.$$
The condition $r_1r_2=e^2$, squared, becomes a quartic equation in $\rho$:
\begin{equation}
\rho^4+4\rho^3\cos\varphi+4\rho^2-e^4=0. \label{eq:quartique}
\end{equation}
Here lies the whole difficulty: for the ellipse, the focal equation is \emph{linear} in $\cos\varphi$ and of degree~1 in $r$ - a conic. Here it remains linear in $\cos\varphi$ - which will prove valuable - but is of degree~4 in $\rho$: Cassini's oval is not a conic, it is a quartic.

Linearity in $\cos\varphi$ nonetheless allows \eqref{eq:quartique} to be inverted explicitly:
\begin{equation}
\cos\varphi = P(\rho) \;\equiv\; \frac{e^4-\rho^4-4\rho^2}{4\rho^3}. \label{eq:Prho}
\end{equation}
In other words, contrary to what one might fear from a quartic, an entirely explicit ``inverse orbit equation'' is available: $\varphi(\rho)=\arccos P(\rho)$ (on the relevant branch). This relation carries the whole of what follows, kinematic and dynamical alike.

The two vertices on the focal axis follow from setting $\cos\varphi=\pm1$ in \eqref{eq:quartique}: $\rho(0)=A-1$ (pericentre, the near vertex) and $\rho(\pi)=A+1$ (apocentre, the far vertex), where
$$A=\sqrt{1+e^2},$$
consistent with the semi-major axis $A$ of the comparison ellipse already introduced in \S\ref{sec:oval-geometry}. Thus $\rho$ ranges, along the whole oval, over the interval $[A-1,A+1]$ - an interval of constant width~2 regardless of $e$, a property with no simple elliptical counterpart.
A conic has genus~0, hence a rational parametrisation and an elementary theory of swept areas. Cassini's oval, by contrast, is a \emph{bicircular quartic} of generic genus~1 (an elliptic curve); only the lemniscate ($e=1$), whose real double point drops the genus to~0, escapes. One therefore expects the swept area to involve elliptic integrals rather than elementary functions and, as \S\ref{sec:eccentric} confirms, no elementary exact Kepler equation.

\paragraph{When the oval is no longer seen once from the focus.} For an ellipse seen from a focus, every direction $\varphi$ meets the curve once, and $\rho(\varphi)$ is a function. For the oval this holds only above a threshold. The map $\rho\mapsto\varphi$ of \eqref{eq:Prho} is monotonic as long as $P'(\rho)\neq0$, where
$$P'(\rho)=-\frac{3e^4}{4\rho^4}-\frac14+\frac1{\rho^2},\qquad P'(\rho)=0\iff\rho^4-4\rho^2+3e^4=0 .$$
This quadratic in $\rho^2$ has real roots iff its discriminant $16-12e^4$ is non-negative, i.e.\ $e\le e_c\equiv(4/3)^{1/4}\approx1.0746$. For $1<e<e_c$ there is a small angular window around $\varphi=\pi$ in which the same direction meets the oval three times, so that $\rho(\varphi)$ is multivalued there; beyond $e_c$ the oval is star-shaped from the focus all along. The threshold is the same as that of the force law (\S\ref{sec:dynamics}), because the same polynomial $-4\rho^4P'(\rho)$ appears in the denominator of the force: the force diverges exactly where the orbit stops being a graph $\rho(\varphi)$. Below $e_c$ the area-law construction used below is therefore not defined by a single-valued polar equation, and all statements about $M(E)$ refer to $e>e_c$.

\subsection{An eccentric anomaly for the oval, and a generalised Kepler equation}
\label{sec:eccentric}

By direct analogy with Kepler, define the mean anomaly on the oval as the fraction of area swept from pericentre:
$$M_{\text{cass}}(\varphi) = 2\pi\,\frac{A(\varphi)}{A_{\text{tot}}}, \qquad A(\varphi)=\frac12\int_0^\varphi \rho(\varphi')^2\,d\varphi', \qquad A_{\text{tot}}=A(2\pi).$$
Since $\rho(\varphi)$ is the root of a quartic and $A(\varphi)$ is in general an elliptic integral, we compute it numerically (an expression through elliptic integrals is possible in principle, but would be of no use for the argument): for each $\varphi$ the quartic \eqref{eq:quartique} is solved for $\rho$, the physical root is selected by continuity, and $\rho^2$ is integrated.

\begin{figure}[H]
\centering
\includegraphics[width=0.62\textwidth]{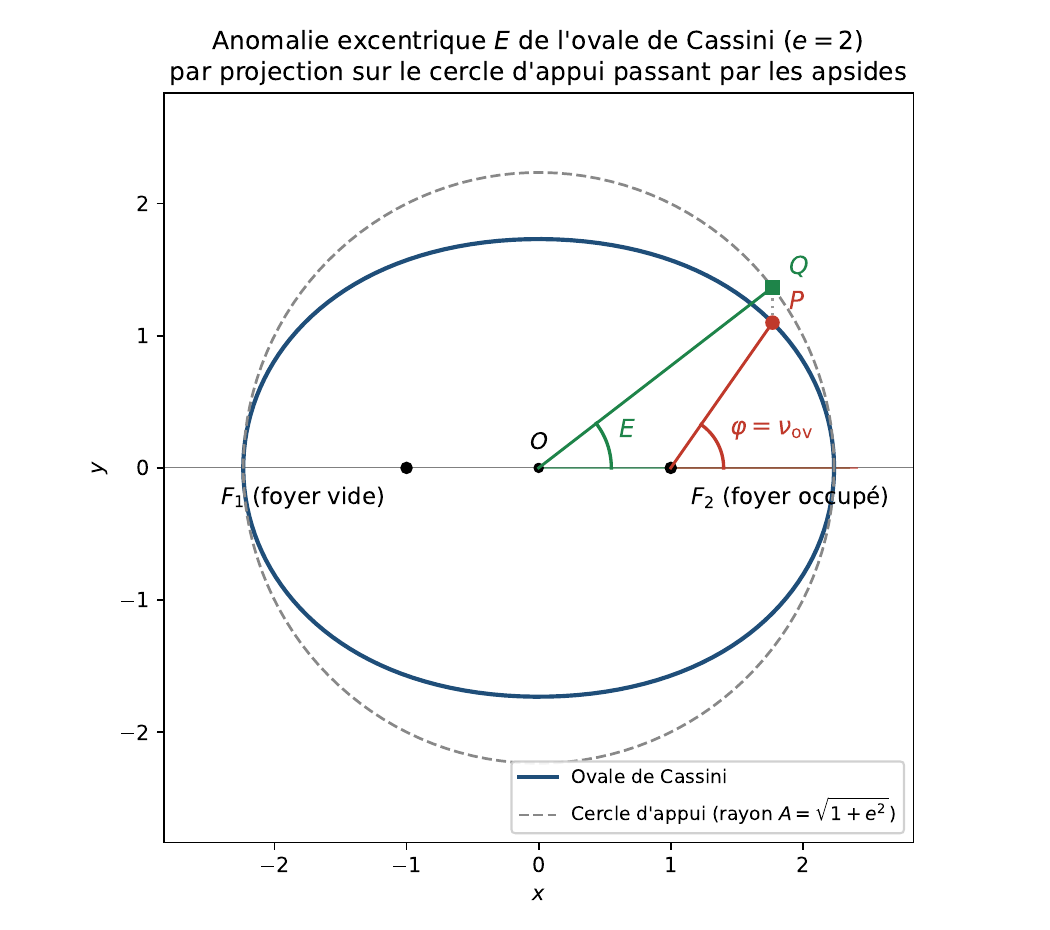}
\caption{Eccentric anomaly $E$ of Cassini's oval ($e=2$), defined by projecting the oval's point $P$ onto the auxiliary circle of radius $A=\sqrt{1+e^2}$ through the two apsides.}
\label{fig:anomalie-excentrique}
\end{figure}

The true anomaly is $\varphi$ itself, and the mean anomaly $M_{\text{cass}}$ has just been defined. What remains is the third element of the classical triad: the \emph{eccentric} anomaly. For an ellipse it is constructed by projecting the orbital point onto the circumscribed circle of radius $a$ centred on the curve's centre - and it is precisely this purely geometric construction, with no reference to dynamics, that lets Ward formulate his equation. Nothing prevents the same construction on the oval, even though it is not a conic: take as auxiliary circle the circle of radius $A=\sqrt{1+e^2}$ centred on the oval's centre $O$ - the circle passing exactly through both apsides, at $(\pm A,0)$ - and define, for a point $P=(x,y)$ on the oval,
\begin{equation}
E \;=\; \arccos\!\Bigl(\frac{x}{A}\Bigr), \qquad \operatorname{sign}(E)=\operatorname{sign}(\varphi), \label{eq:def-E}
\end{equation}
exactly as $x=a\cos E$ for the ellipse (Figure~\ref{fig:anomalie-excentrique}).

Computing the relation $M(E)$ numerically (with $M=M_{\text{cass}}$) over a dense grid of $e$ and decomposing $E-M$ as a Fourier series in $E$, $E-M=\sum_{k\ge1}b_k(e)\sin(kE)$, gives a striking result: for every value of $e$ tested ($1.42\le e\le200$), to machine precision,
\begin{equation}
b_1(e) \;=\; \frac{1}{A} \;=\; \frac{1}{\sqrt{1+e^2}}, \label{eq:b1-exact}
\end{equation}
which is exactly the analogue of Kepler's own coefficient $\varepsilon=c/a$ (focal distance over semi-major axis), with $c=1$ here and $A$ playing the role of the semi-major axis. This is not a numerical coincidence: it follows from the quartic~\eqref{eq:quartique} itself, by an exact argument given in Appendix~\ref{sec:appendix-proof}; it was also confirmed numerically, to $10^{-13}$--$10^{-15}$, for every value of $e$ tested.

\begin{figure}[H]
\centering
\includegraphics[width=0.92\textwidth]{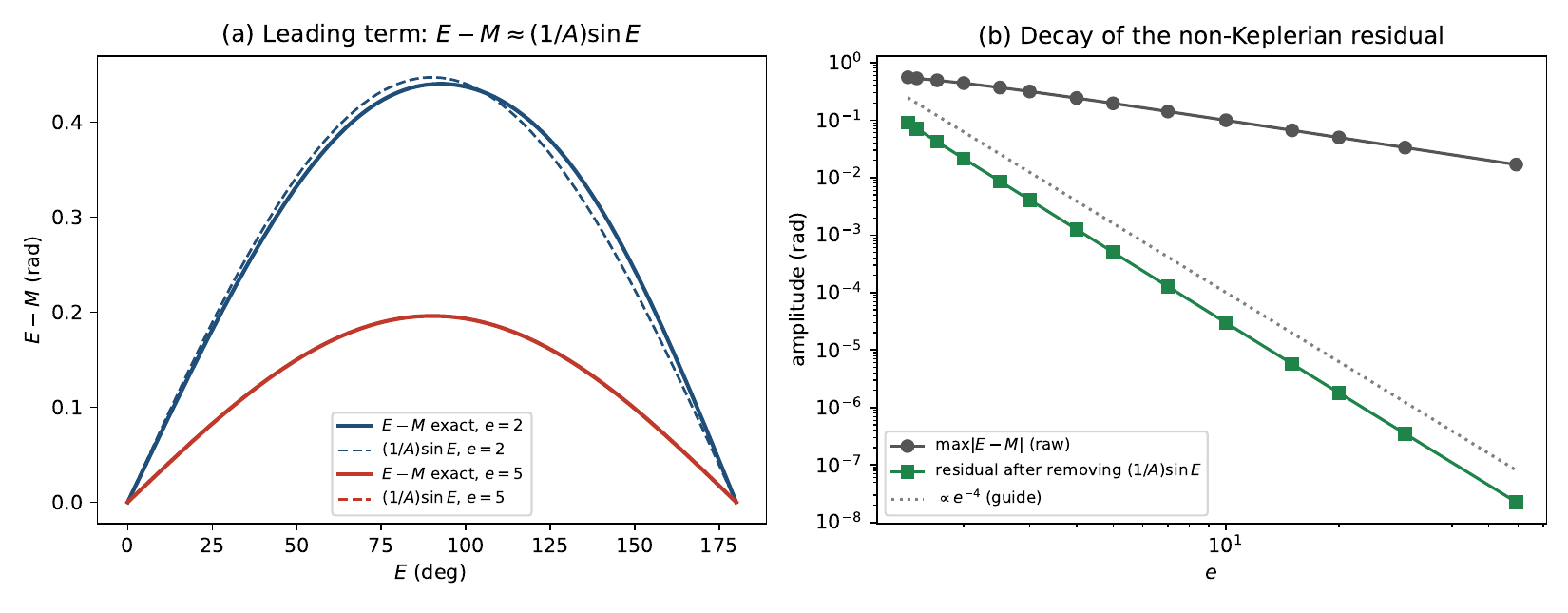}
\caption{(a) The discrepancy $E-M$ (solid) compared with its exact leading term $(1/A)\sin E$ (dashed), for $e=2$ and $e=5$. (b) Maximum amplitude of $E-M$ before (circles) and after (squares) removing the leading term, as a function of $e$: the residual - the signature of the curve's genus~1 - decays much faster than the term itself.}
\label{fig:kepler-like}
\end{figure}

Unlike the ellipse - where Kepler's equation $M=E-\varepsilon\sin E$ is \emph{exact}, with no higher-order term at all - the subsequent harmonics $b_2,b_3,\ldots$ do \emph{not} vanish for the oval: this is the direct signature of its genus-1 nature (\S\ref{sec:quartic}), rather than genus~0 as for a conic. There is, accordingly, no elementary exact Kepler equation for Cassini's oval - an expected result, since such an equation, a finite sum of sine terms linking $E$ and $M$, would amount to a rational parametrisation of the curve, excluded by its genus (except at the degenerate lemniscate, $e=1$, whose genus is~0). But the truncated approximation
\begin{equation}
M \;\approx\; E - \frac{1}{A}\sin E \label{eq:kepler-generalisee}
\end{equation}
proves remarkably accurate: Figure~\ref{fig:kepler-like}(a) compares it with the exact relation for $e=2$ and $e=5$; panel~(b) shows that the residual after removing this leading term already falls to $5\times10^{-4}$~rad by $e=5$ (Mercury) - some $390$ times below the raw discrepancy $M-E$ itself - and to about $2\times10^{-8}$~rad for the Sun ($e\approx59.3$).

A full symbolic expansion in $\varepsilon=1/e$ (Appendix~\ref{sec:appendix}), carried to fifth order, confirms and refines this: $b_2(e)\sim-\varepsilon^4/4$, $b_3(e)\sim\varepsilon^5/4$, a previously unnoticed fourth harmonic $b_4(e)\sim-\varepsilon^4/16$ of the same order as $b_2$, and a vanishing leading term for $b_5$ (which only appears at $O(\varepsilon^6)$). The resulting fifth-order formula
\begin{align*}
M \;&=\; E \;-\; \frac1A\sin E \;+\; \frac{\varepsilon^4}{4}\sin2E \\ &\quad \;-\; \frac{\varepsilon^5}{4}\sin3E \;+\; \frac{\varepsilon^4}{16}\sin4E \;+\; O(\varepsilon^6)
\end{align*}
was obtained by two entirely independent routes - a global integral identity and a local, order-by-order expansion of the defining quartic - that agree on the same coefficients $1,-\tfrac12,\tfrac38$ in the Taylor expansion of $1/A$, an independent cross-check of \eqref{eq:b1-exact}. The full derivation, term by term, is given in Appendix~\ref{sec:appendix}.

In short, Cassini's oval is neither a conic nor a Keplerian curve in the strict sense - nothing in its quartic equation would have suggested otherwise - yet it shares, unexpectedly and to a precision sufficient for astronomical use, several of the kinematic properties that make Kepler's equation practically useful: a near-equant empty focus (\S\ref{sec:equant}), and an eccentric-anomaly relation whose leading term coincides, coefficient for coefficient, with the ellipse's own. Section~\ref{sec:reception} asks whether Cassini, Ward, or Jacques Cassini could, in principle, have noticed any of this.
\subsection{Jacques Cassini's 1740 calculation: the one worked example}
\label{sec:jc-example}

The only calculation made by an astronomer from the oval is Jacques Cassini's, in the \emph{\'El\'ements d'astronomie} (1740), Book~II, chap.~vi, pp.~149--151: his father, he says, found a curve other than the ellipse that ``serves to represent very exactly the true motions of the Sun'', the product of the distances to the two foci being equal to the product of the Sun's greatest and least distances. With the Earth at one focus, the construction is that of an equant: the angle at the \emph{empty} focus, taken as proportional to time, plays the role of the mean anomaly. We call it $M_{\mathrm{eq}}$ to distinguish it from the area-defined mean anomaly $M_{\mathrm{aire}}$ defined above and used in \S\ref{sec:equant}. These are two different time laws on the same curve; Jacques Cassini uses only the first, and nothing in the passages quoted attributes an area law to the Cassinis.

\begin{figure}[H]
\centering
\includegraphics[width=0.72\textwidth]{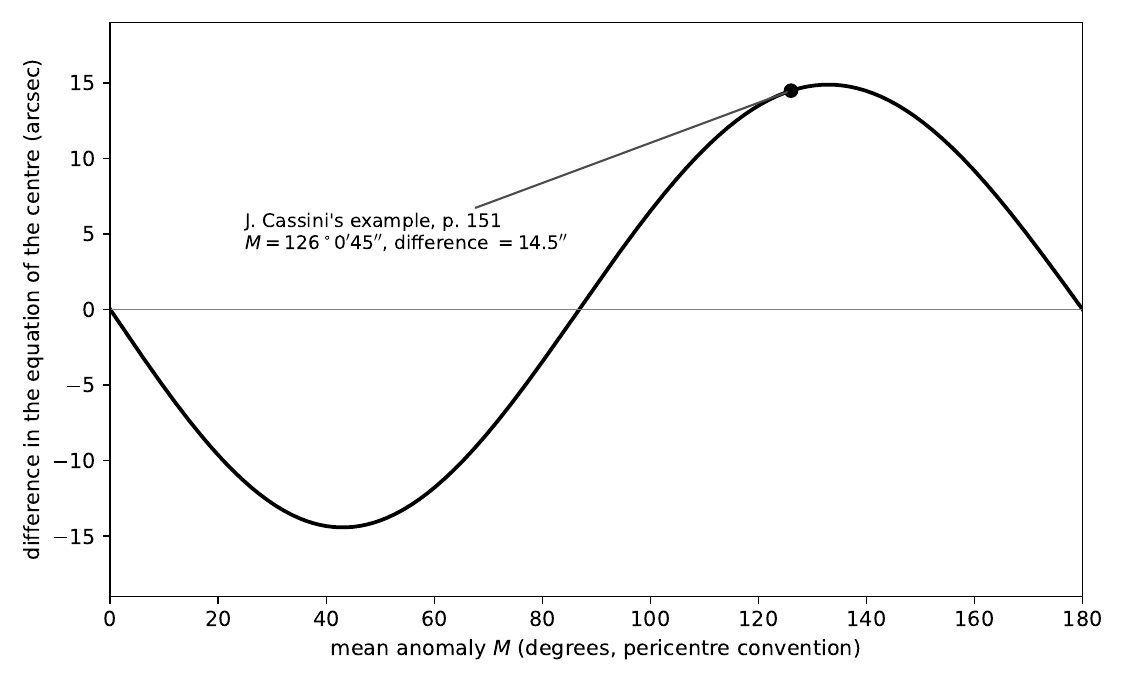}
\caption{Difference between the equations of the centre of the oval (with equant at the empty focus) and of Kepler's ellipse, for Jacques Cassini's Sun ($\varepsilon=0.01685$), as a function of mean anomaly (pericentre convention). The marked point is the worked example of p.~151 ($M=126^\circ0'45''$, $14.5''$).}
\label{fig:ecart_M}
\end{figure}

Appendix~\ref{sec:appendix-jc} reconstructs the calculation from his numbers (aphelion $1\,016\,850$, perihelion $983\,150$, hence $\varepsilon=0.01685$, and mean distance $10^6$). For the Earth--Sun distance $FH=1\,010\,000$ the oval gives $HE=989\,818$, his value; the triangle $EFH$ gives an equation of the centre $\nu_{\mathrm{ov}}-M_{\mathrm{eq}}=1^\circ32'47''$ at $M_{\mathrm{eq}}=126^\circ0'45''$, against $1^\circ32'33''$ for Kepler's ellipse at the same mean anomaly: a difference of $14''$, which he judges ``scarcely perceptible in observations.'' Over the whole orbit (Fig.~\ref{fig:ecart_M}), the difference between the oval with equant and the Keplerian ellipse is nearly sinusoidal, vanishing at the apsides, with a maximum of $14.89''$ near $M=133^\circ$. His chosen point ($14.5''$) lies within $7^\circ$ of this maximum; that he picked it for that reason is a possibility, not something the text establishes. The leading term is $\mp(\varepsilon^2/4)\sin2M$ ($14.64''$ for the Sun), but the series is only a guide: the exact curve differs from it by up to $1.6''$, and is only approximately antisymmetric about $90^\circ$.

This $14.89''$ compares an \emph{oval with equant} to a \emph{Keplerian ellipse}. It is not the $14.88''$ of \S\ref{sec:equant}, which compares the ellipse's own empty-focus angle to its Keplerian mean anomaly; the near-equality of the two is itself a result, since it means that the oval, with an equant, reproduces a Kepler ellipse to nearly the accuracy with which an equant on the ellipse itself does. For the Sun, $e=\sqrt{1-\varepsilon^2}/\varepsilon\approx59.3$ and $\mathcal D(e)\simeq2\times10^{-4}$.

\section{Dynamics: the force law via Binet's formula}
\label{sec:dynamics}

Suppose now - reversing Kepler's own procedure - that a particle genuinely traces the Cassini oval under a central force $F(\rho)$, with conserved angular momentum $L=\rho^2\dot\varphi$ (the area law). Binet's formula expresses the required force purely from the orbit's shape, $u(\varphi)=1/\rho(\varphi)$:
$$F(\rho) = -mL^2u^2\left(\frac{d^2u}{d\varphi^2}+u\right).$$
Writing $f(\rho)=u^2(d^2u/d\varphi^2+u)$, so that $F=-mL^2f(\rho)$ (chosen so that $f>0$ is attractive, and the Keplerian case $F\propto1/\rho^2$ gives $f(\rho)=GM/L^2=$ const), define the natural indicator
$$g(\rho) \equiv \rho^2 f(\rho),$$
which is \emph{constant} iff the force is exactly $1/\rho^2$ - the Newtonian criterion that distinguishes the inverse square among power laws.

\paragraph{An exact result, without elliptic integrals.} Remarkably, $g(\rho)$ is available in closed form, despite the absence of one for $\rho(\varphi)$ itself, because \eqref{eq:Prho} lets $\rho'\equiv d\rho/d\varphi$ be written as a function of $\rho$ alone, from which $f(\rho)$, and hence $g(\rho)=\rho^2f(\rho)$, follow by the chain rule - a symbolic computation (\emph{sympy}) that collapses entirely (all square roots cancel) into the rational fraction
\begin{equation}
g(\rho,e) = \frac{8e^4\bigl(3e^8+2e^4\rho^4-12e^4\rho^2+3\rho^8-12\rho^6+16\rho^4\bigr)}{\rho\,(3e^4+\rho^4-4\rho^2)^3}. \label{eq:gform}
\end{equation}
The polynomial $3e^4+\rho^4-4\rho^2$ in the denominator is $-4\rho^4P'(\rho)$, with $P$ from \eqref{eq:Prho}. Since $\cos\varphi=P(\rho)$, one has $\rho'=d\rho/d\varphi=-\sin\varphi/P'(\rho)$, and each further derivative with respect to $\varphi$ brings another factor $1/P'(\rho)$; the three powers that result in $f$, which involves $\rho''$, are exactly the cube in \eqref{eq:gform}. The required force diverges where this polynomial vanishes.

\paragraph{Three regimes.}\label{sec:regimes} Analysing \eqref{eq:gform} on the physical interval $\rho\in[A-1,A+1]$ gives three cases, matching the geometric ones of \S\ref{sec:oval-geometry}. For $1<e<(4/3)^{1/4}\approx1.0746$ the denominator vanishes inside the interval and the force diverges: no finite central force can produce this motion. For $(4/3)^{1/4}<e<\sqrt2$ the force is finite but changes sign along the orbit, alternately attractive and repulsive; this range coincides exactly with the non-convex ``dumbbell'' silhouettes, an independent cross-check between a geometric criterion (convexity) and a dynamical one. At $e=\sqrt2$, the limit of convexity, the force is non-negative and touches zero at the single radius $\rho=\sqrt2$ without changing sign, so that $g_{\min}=0$ and $\mathcal D(\sqrt2)=1$ (the index reaches its maximum there, at the boundary between the two regimes). For $e>\sqrt2$, the strictly convex ovals, the force is everywhere finite and attractive, but not Keplerian. This last regime is the only one in which comparison with $1/\rho^2$ is well posed, and the one relevant to the planets ($e\approx59$ for the Sun, $4.8$ for Mercury).

\paragraph{Quantifying the departure from the inverse square.}\label{sec:deviation}

\begin{figure}[H]
\centering
\includegraphics[width=0.72\textwidth]{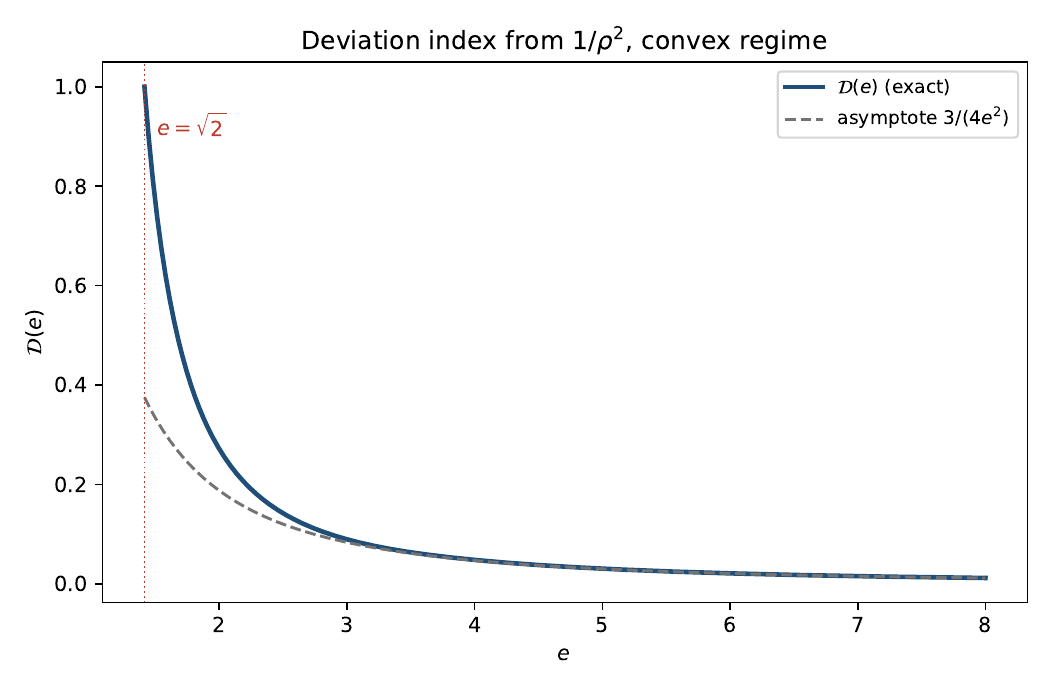}
\caption{Deviation index $\mathcal{D}(e)$ from the $1/\rho^2$ law required to produce a Cassini oval of parameter $e$ under the area law, in the convex regime $e\geq\sqrt2$. The dashed curve is the asymptote $3/(4e^2)$.}
\label{fig:deviation}
\end{figure}

Bertrand's theorem states that, among central forces, only $F\propto1/\rho^2$ and $F\propto\rho$ produce closed orbits for every initial condition. It does not forbid a central force from producing one particular oval-shaped closed orbit - formula \eqref{eq:gform} constructs such a force - and it says nothing about how far that force departs from the inverse square. For the family of Cassini ovals this can be answered by a simple numerical index. In regime~III define
$$\mathcal{D}(e) \;\equiv\; \frac{g_{\max}-g_{\min}}{g_{\max}+g_{\min}}, \qquad g_{\max,\min}=\max_{\rho\in[A-1,A+1]}\!,\min\, g(\rho,e),$$
which is $0$ for an exact inverse-square force and tends to $1$ as the force departs strongly from it. Figure~\ref{fig:deviation} plots it for $e\geq\sqrt2$, and Table~\ref{tab:D} gives selected values.

\begin{table}[H]
\centering
\begin{tabular}{@{}lcccc@{}}
\toprule
$e$ & $\sqrt2\approx1.41421$ & $1.5$ & $2$ & $5$ \\
\midrule
$\mathcal{D}(e)$ & $1.0000$ & $0.7668$ & $0.2713$ & $0.0305$ \\
\bottomrule
\end{tabular}
\caption{Deviation index $\mathcal{D}(e)$ from the inverse square, selected values.}
\label{tab:D}
\end{table}

A series expansion (substitution $e=1/\varepsilon$, $\rho=A+s$, expanding \eqref{eq:gform} to fourth order as $\varepsilon\to0$) gives $g(\rho,e)\approx\varepsilon+\tfrac32\varepsilon^3s^2+O(\varepsilon^4)$, from which
\begin{equation}
\mathcal{D}(e) \;\sim\; \frac{3}{4e^2} \qquad (e\to\infty), \label{eq:asympt}
\end{equation}
verified numerically to high precision (fitted constant $\approx0.7525$ against the expected $0.75$). This is the quantitative answer Bertrand's qualitative theorem does not supply: the force law required to produce a Cassini oval of parameter $e$ departs from the inverse square by a relative amount decreasing as $3/(4e^2)$ - not, say, $1/e$ or $1/e^4$. This index measures a departure; by itself it does not prove that no conceivable law could serve the oval family, only that the force that does so depends on $e$ and is never exactly inverse-square.

\section{An (almost) equant focus}
\label{sec:equant}

Cassini himself, in his 1693 \emph{Traité}, remarked that a planet's real speed increases as it nears the Sun and decreases as it recedes (\S\ref{sec:reception}) - so he must, at some level, have registered that speed along his own curve is not uniform when viewed from the occupied focus. This raises a natural question, purely kinematic and independent of \S\ref{sec:dynamics}: viewed from \emph{which} focus, if either, does motion along the oval come closest to uniform - and could the empty focus play something like the role of a Ptolemaic equant, exactly as Ward's construction proposes for the ellipse?

Consider a particle moving along the oval such that the true area law $L=\rho^2\dot\varphi$ holds about the occupied focus $F_2$ (the dynamically motivated convention of \S\ref{sec:dynamics}). Writing each point as a complex number relative to a chosen pole and using $\dot\theta=\operatorname{Im}(\dot z/z)$, a symbolic computation gives the polar angular velocity about the \emph{empty} focus $F_1$ in closed form,
$$\dot\theta_{F_1} \;=\; L\cdot h(\rho,e), \qquad h(\rho,e)=\frac{e^4+3\rho^4-4\rho^2}{\rho^2(3e^4+\rho^4-4\rho^2)},$$
whose asymptotic expansion for large $e$ (writing $\rho=A+s$, $\varepsilon=1/e$) is
$$h(A+s,e)=\varepsilon^2+2s(1-s^2)\varepsilon^5+(3s^4-4s^2+1)\varepsilon^6+O(\varepsilon^7).$$
The leading term is \emph{exactly constant}, with no angular dependence at all before fifth order: the empty focus is, to a striking degree, a quasi-equant. Defining a deviation index $\mathcal{D}_\theta(e)$ on the same pattern as $\mathcal{D}(e)$ gives the exact extremum of $2s(1-s^2)$ at $s=\pm1/\sqrt3$ and hence
$$\mathcal{D}_\theta(e)\;\sim\;\frac{4\sqrt3}{9e^3} \qquad (e\to\infty),$$
an $O(\varepsilon^3)$ relative deviation - two orders later, in the small parameter, than the naive equant approximation would suggest. A parallel computation for the \emph{other} direction (uniform motion assumed at the empty focus) gives $k(\rho,e)=1/h(\rho,e)$ exactly: the two foci are, to this order, dynamically interchangeable, a symmetric ``pseudo area law'' linking them.

\subsection{Comparison with Ward's own ellipse}
\label{sec:ellipse-vs-oval}

The natural next question is whether this near-uniformity is in any sense special to the oval, or whether the ellipse's own empty focus - the very point for which Ward devised the construction - does at least as well. The same calculation, run on the real ellipse under the true Keplerian area law ($\dot\nu=L/r^2$), gives
\begin{gather*}\dot\theta_E \;=\; L\cdot h_{\text{ell}}(\nu,\varepsilon), \\ h_{\text{ell}}(\nu,\varepsilon) = \frac{(1+\varepsilon\cos\nu)^2(1+2\varepsilon\cos\nu+\varepsilon^2)}{(1-\varepsilon^2)(1+4\varepsilon\cos\nu+2\varepsilon^2+4\varepsilon^3\cos\nu+4\varepsilon^2\cos^2\nu+\varepsilon^4)},\end{gather*}
whose small-$\varepsilon$ expansion, $h_{\text{ell}}=1+\varepsilon^2\cos^2\nu+O(\varepsilon^3)$, carries a non-vanishing angular dependence already at second order - unlike the oval, whose leading term had none. The corresponding deviation index is exact, $\mathcal{D}_{\theta,\text{ell}}(\varepsilon)=\varepsilon^2/(2-\varepsilon^2)\sim\varepsilon^2/2$, and the resulting maximum angular discrepancy is
$$\theta_E(M)-M \;=\; \frac{\varepsilon^2}{4}\sin2M \;+\; O(\varepsilon^3),$$
of amplitude $\varepsilon^2/4$, i.e.\ $14.64''$ for the Sun's eccentricity (series prediction), against a numerically exact maximum of $14.88''$ - exactly the $e^2/16$ figure already met, in Heilbron's differing normalisation, in \S\ref{sec:intro}: once the two conventions are reconciled ($e=2\varepsilon$ there), $(2\varepsilon)^2/16=\varepsilon^2/4$, and the two results coincide exactly.

\begin{table}[H]
\centering
\begin{tabular}{@{}lccc@{}}
\toprule
& $\varepsilon$ (ellipse) vs $e$ (oval) & $\mathcal{D}_{\theta,\text{ell}}(\varepsilon)$ & $\mathcal{D}_\theta(e)$ (oval) \\
\midrule
Sun & $0.01685$ ($e\approx59.3$) & $1.420\times10^{-4}$ & $3.69\times10^{-6}$ \\
Mercury & $0.206$ ($e\approx4.76$) & $2.168\times10^{-2}$ & $7.30\times10^{-3}$ \\
\bottomrule
\end{tabular}
\caption{Deviation from uniform polar velocity at the empty focus, ellipse vs.\ oval, at the same two classical eccentricities used throughout.}
\label{tab:comparaison-ellipse-oval}
\end{table}

For the Sun, the oval's empty focus is roughly \textbf{38 times} closer to uniformity than the ellipse's own empty focus; for Mercury the advantage narrows but remains a factor of about~3. Across the whole range of solar-system eccentricities, the equant device turns out to be \emph{structurally better suited to Cassini's oval than to the very ellipse for which Ward invented it} - a full order of magnitude ($\varepsilon^3$ against $\varepsilon^2$) separates the two asymptotic laws.

This is, at first sight, paradoxical: the less ``Keplerian'' of the two curves - the one admitting neither an exact Kepler equation nor an inverse-square force law - turns out to suit \emph{better} the very device Ward invented for the other. The explanation lies in a structural fact already visible above: for the oval, the leading term of $h(\rho,e)$ is \emph{exactly} constant, with no angular dependence before fifth order; for the ellipse, angular dependence ($\cos^2\nu$) appears at the very first non-trivial order. This early cancellation, specific to the oval's bifocal geometry and nowhere predictable \emph{a priori}, accounts for the whole gap between the two columns of Table~\ref{tab:comparaison-ellipse-oval}.

\paragraph{Summary of the comparison.} Table~\ref{tab:summary} collects the point-by-point verdict developed across \S\S\ref{sec:kinematics}--\ref{sec:ellipse-vs-oval}. The oval matches or beats Ward's equant model on every kinematic count examined - the pseudo-Kepler-equation's leading coefficient is exact, and the empty focus is a better equant by more than an order of magnitude - but this advantage does not carry over to a genuine central force: the force required by the oval depends on $e$ (\S\ref{sec:deviation}).

\begin{table}[H]
\centering
\begin{tabular}{@{}p{0.27\textwidth}p{0.32\textwidth}p{0.33\textwidth}@{}}
\toprule
Criterion & Ellipse (area law) & Cassini's oval \\
\midrule
Leading eccentric-anomaly coeff.\ & exact, for the area law ($M=E-\varepsilon\sin E$) & exact, identical form (eq.~\ref{eq:b1-exact}) \\
Empty-focus deviation index & $\sim\varepsilon^2/2$ & $\sim4\sqrt3/(9e^3)$, $\sim38\times$ better (Sun) \\
Exact central force (area law) & $1/\rho^2$, always & none universal; diverges/changes sign below $\sqrt2$ \\
\bottomrule
\end{tabular}
\caption{Point-by-point comparison of the two rival responses to the difficulty of Kepler's equation.}
\label{tab:summary}
\end{table}

\section{Universality, and why the question stayed open}
\label{sec:history}
\label{sec:newton}

Sections~\ref{sec:kinematics}--\ref{sec:equant} have shown a curve that is kinematically close to Kepler's (same leading coefficient, near-equant empty focus) but dynamically far from it. We return to the two questions of Section~\ref{sec:reception}, and state precisely what the argument does and does not claim. Newton's first proposition in the \emph{Principia} - that an observed area law, about any pole, signs a central force - is indifferent to the shape of the orbit; nothing would have stopped Newton from running that argument on Cassini's oval, and \S\ref{sec:kinematics} shows the resulting kinematics to be perfectly well posed. What breaks down is the next step, the one that carries Book~I into Book~III: extracting, from the orbit's curvature, a force law that is not merely central but \emph{universal} - the same power law, with the same exponent, for every planet. Section~\ref{sec:deviation}'s index $\mathcal{D}(e)\sim3/(4e^2)$ shows that no such universal law exists for the oval family: the required force depends on $e$ in a way that is not even always of constant sign (regimes~I and~II, \S\ref{sec:regimes}), and approaches $1/\rho^2$ only asymptotically, planet by planet, never exactly. Kepler's ellipse, by contrast, yields $1/\rho^2$ for every eccentricity $\varepsilon$ without exception. It is this absence of a single, universal exponent - not the absence of an area law as such - that, we suggest, made the oval unfit for the programme completed in Book~III. The letters cited in \S\ref{sec:reception} document uncertainty, not this causal chain: it is an interpretation consistent with the evidence, not a documented reason.

Formulated this way, the counterfactual is no longer a matter of plausible intuition but a direct consequence, quantified above, of the oval's own dynamics. It also helps explain why competent contemporaries did not treat the matter as settled quickly. Heilbron reports that although ``later astronomers made fun of these curves,'' they in fact ``earned Cassini compliments from Newton and [\ldots] were preferred, for a time, to Kepler's, in Paris''\footnote{Heilbron, \emph{Astronomie et églises} (ref.~\nr{n2}), chap.~3, pp.~118-19.} - the retrospective mockery is a later verdict, not that of Newton's own contemporaries. Cohen supplies a precise, dated institutional testimony: the \emph{Histoire de l'Académie royale des sciences} for the year 1703 - some twenty years after Cassini advanced his oval, and sixteen years after the \emph{Principia} - states, of the curve Cassini had substituted for Kepler's ellipse, that ``this Ellipse answers better to the celestial observations, if indeed any regular Curve answers to them exactly''\footnote{\emph{Histoire de l'Académie royale des sciences}, year 1703 (Paris, 1720), p.~68, cited in Cohen, ``Leibniz on elliptical orbits'' (ref.~\nr{n16}), 79-80.}. Coming from the institution itself, not from Cassini alone, this judgment does not contradict the thesis developed above - it reinforces it: if the Academy could still, in 1703, seriously entertain the possibility that a non-Keplerian curve ``answers better to the observations,'' the area law plainly did not yet carry, at that date, the status of a decisive criterion it would later acquire. The same passage notes that the marquis de L'Hôpital found a general rule for the tangent to this type of bifocal curve, and that Varignon applied it to Cassini's oval: the curve was, at the turn of the eighteenth century, the object of active technical interest among the Academy's foremost geometers, not of principled rejection.

Given the numbers established in \S\ref{sec:deviation} and \S\ref{sec:ellipse-vs-oval} - discrepancies of order $14''$ for the Sun, a few arcminutes at most for Mercury, and a force-law deviation $\mathcal{D}(e)\simeq2\times10^{-4}$ for solar eccentricities - this institutional hesitation looks considerably less like negligence than like a reasonable response to numbers sitting at, or just beyond, the edge of what late-seventeenth-century instruments could resolve.

\section{Conclusion}
\label{sec:conclusion}

Cassini's oval occupies a singular position in the history of seventeenth-century astronomy: kinematically, it is almost indistinguishable from Kepler's ellipse, sharing the leading coefficient of its eccentric anomaly and offering, at its empty focus, a better equant than the ellipse's own; dynamically, the force it requires departs from the inverse square by a measurable but modest amount - an index $\mathcal{D}(e)\sim3/(4e^2)$, of order $2\times10^{-4}$ for the Sun - yet never vanishes exactly, and changes sign or diverges altogether below $e=\sqrt2$. This article has argued that the second fact, not the first, is what matters historically. Our reading is that the oval was set aside not for want of an area law - that law can be imposed kinematically, and remains dynamically viable outside a narrow divergent regime - but for want of a universal force law, the precondition for Book~III of the \emph{Principia}. We stress that this is a conjecture, drawn both from our calculations and from the reactions of contemporary scientists; we have found no text of the period that states it.

This reading cuts against two historiographical commonplaces. The first is that Kepler's laws were rapidly accepted after 1609; the correspondence of Boulliau, Ward, and Leibniz suggests, rather, that they were not, and that the ellipse remained a live question at the Académie as late as 1703. The second is that Newton's \emph{Principia} (1687) closed the debate at once; the evidence assembled in \S\ref{sec:history} suggests, as we read it, that the decisive criterion - universality of the force law - took a generation to be absorbed into astronomical practice. Cassini's oval is interesting precisely because it sits at the intersection of these two commonplaces: close enough to the ellipse to resist refutation by observation, far enough from it to resist absorption into the Newtonian programme, it remained, for a generation, an option that only the gradual triumph of universality could render obsolete.

\theendnotes

\appendix
\section{Proof of equation~\eqref{eq:b1-exact}}
\label{sec:appendix-proof}

Throughout, $e>(4/3)^{1/4}$, so that the oval is star-shaped from its centre $O$ and from the focus $F_2$; let $D$ be the region it bounds. The area swept from $F_2$ over the whole orbit is $\mathcal S=\int_0^{2\pi}\tfrac12\rho^2\,d\varphi=\operatorname{area}(D)$, and by symmetry $\int_0^{\pi}\rho^2\,d\varphi=\mathcal S$.

\paragraph{Step 1: reduction.} On $[0,\pi]$ the function $E-M$ vanishes at both ends, and its sine coefficients are $b_k=\tfrac2\pi\int_0^\pi(E-M)\sin kE\,dE$. Integrating by parts for $k=1$, the boundary terms vanish and $\int_0^\pi\cos E\,dE=0$, so
$$b_1=-\frac2\pi\int_0^\pi\cos E\,dM .$$
Now $\cos E=x/A=(1+\rho\cos\varphi)/A$ (abscissa measured from $O$) and $dM=\pi\rho^2\,d\varphi/\mathcal S$, whence
\begin{gather*}b_1=-\frac{2}{A\mathcal S}\int_0^\pi(\rho^2+\rho^3\cos\varphi)\,d\varphi=-\frac1A\Bigl(2+\frac{I_3}{\mathcal S}\Bigr),\\ I_3=\int_0^{2\pi}\rho^3\cos\varphi\,d\varphi .\end{gather*}
Equation~\eqref{eq:b1-exact} is therefore equivalent to $I_3=-3\mathcal S$.

\paragraph{Step 2: the identity $I_3=-3\mathcal S$.} The quartic~\eqref{eq:quartique} gives $\rho^3\cos\varphi=(e^4-\rho^4-4\rho^2)/4$, hence
$$I_3=\tfrac14\Bigl[2\pi e^4-\int_0^{2\pi}\rho^4\,d\varphi-8\mathcal S\Bigr],$$
so that $I_3=-3\mathcal S$ is equivalent to $\int_0^{2\pi}\rho^4\,d\varphi=2\pi e^4+4\mathcal S$. Since $D$ is star-shaped from $F_2$, $\int_0^{2\pi}\rho^4\,d\varphi=4\iint_D\rho^2\,dA$. With $z=x+iy$ measured from $O$, $\rho^2=|z|^2-2x+1$ and $\iint_Dx\,dA=0$ by symmetry, so that
$$\int_0^{2\pi}\rho^4\,d\varphi=4J+4\mathcal S,\qquad J=\iint_D|z|^2\,dA=\tfrac14\int_0^{2\pi}r^4\,d\theta,$$
$J$ being the polar moment of $D$ about $O$. The centred quartic gives $r^4=2r^2\cos2\theta+e^4-1$, and with $r^2=\cos2\theta+\sqrt{e^4-\sin^22\theta}$ one finds $\int_0^{2\pi}r^2\cos2\theta\,d\theta=\pi$, the square-root term integrating to zero under $\theta\mapsto\pi/2-\theta$. Hence $J=\pi e^4/2$, the required identity $\int_0^{2\pi}\rho^4\,d\varphi=2\pi e^4+4\mathcal S$ holds, and $b_1=1/A$ exactly. \qed

\section{The fifth-order expansion of the generalised Kepler equation}
\label{sec:appendix}

This appendix gives the full derivation, by computer algebra, of the fifth-order expansion in $\varepsilon=1/e$ announced in \S\ref{sec:eccentric}. The method follows the pattern already used in \S\ref{sec:equant}: substitute $\rho=A(1+\varepsilon v(\varphi,\varepsilon))$ into the defining quartic~\eqref{eq:quartique} and solve order by order in $\varepsilon$ for $v=v_0+v_1\varepsilon+\cdots+v_5\varepsilon^5$, giving
\begin{gather*}
v_0=-\cos\varphi, \quad v_1=-\tfrac32\sin^2\varphi,\\
v_2=\cos\varphi-\tfrac12\cos3\varphi, \quad v_3=\tfrac18(15\sin^2\varphi-4)\sin^2\varphi,
\end{gather*}
\begin{gather*}
v_4=-3\sin^4\varphi\cos\varphi-\tfrac14\cos\varphi-\tfrac5{16}\cos3\varphi+\tfrac3{16}\cos5\varphi,\\ v_5=\tfrac{45}{16}\sin^6\varphi-\tfrac{11}4\sin^4\varphi-\tfrac58\sin^2\varphi+\tfrac18\sin^23\varphi.
\end{gather*}

From $\cos E=x/A=1/A+(1+\varepsilon v)\cos\varphi$, writing $E=\varphi+\delta(\varphi,\varepsilon)$ and solving order by order gives
\begin{gather*}
\delta_1=-\sin\varphi, \quad \delta_2=\tfrac12\sin2\varphi,\\
\delta_3=\tfrac18\sin\varphi-\tfrac5{24}\sin3\varphi, \quad \delta_4=-\tfrac14\sin2\varphi,
\end{gather*}
$$\delta_5=-\tfrac{15}{64}\sin\varphi+\tfrac{11}{128}\sin3\varphi+\tfrac{77}{640}\sin5\varphi.$$

Separately, $M(\varphi,\varepsilon)=\varphi+\mu(\varphi,\varepsilon)$ follows from the ratio of swept areas, $M=\pi\int_0^\varphi(1+\varepsilon v)^2\,d\varphi'\big/\int_0^\pi(1+\varepsilon v)^2\,d\varphi'$, giving
\begin{gather*}
\mu_1=-2\sin\varphi, \quad \mu_2=\sin2\varphi,\\
\mu_3=\tfrac34\sin\varphi-\tfrac7{12}\sin3\varphi, \quad \mu_4=-\tfrac12\sin2\varphi+\tfrac5{16}\sin4\varphi,
\end{gather*}
$$\mu_5=-\tfrac{39}{20}\sin^5\varphi+\tfrac78\sin\varphi-\tfrac38\sin3\varphi.$$

Reverting the series $E=\varphi+\delta(\varphi,\varepsilon)$ to express $\varphi$ as a function of $E$,
\begin{align*}\varphi&=E+\sin E\,\varepsilon+0\cdot\varepsilon^2-\tfrac16\sin3E\,\varepsilon^3\\ &\quad +\tfrac14\sin2E\,\varepsilon^4+\Bigl(\tfrac18\sin E+\tfrac18\sin3E+\tfrac3{40}\sin5E\Bigr)\varepsilon^5+O(\varepsilon^6),\end{align*}
and substituting into $E-M=\delta(\varphi,\varepsilon)-\mu(\varphi,\varepsilon)$, expanded to fifth order about $\varphi=E$, gives
\begin{align*}E-M&=\Bigl(\varepsilon-\tfrac12\varepsilon^3+\tfrac38\varepsilon^5\Bigr)\sin E\\ &\quad -\Bigl(\tfrac14\sin2E+\tfrac1{16}\sin4E\Bigr)\varepsilon^4+\tfrac14\sin3E\,\varepsilon^5+O(\varepsilon^6),\end{align*}
whence, since $\varepsilon-\tfrac12\varepsilon^3+\tfrac38\varepsilon^5$ is exactly the Taylor expansion of $1/A=\varepsilon/\sqrt{1+\varepsilon^2}$ to this order - an independent confirmation of the exact identity~\eqref{eq:b1-exact}, obtained here by a purely local, order-by-order route rather than the global integral identity of \S\ref{sec:eccentric} - the generalised Kepler equation quoted in \S\ref{sec:eccentric} follows:
\begin{align*}
M \;&=\; E \;-\; \frac1A\sin E \;+\; \frac{\varepsilon^4}{4}\sin2E \\ &\quad \;-\; \frac{\varepsilon^5}{4}\sin3E \;+\; \frac{\varepsilon^4}{16}\sin4E \;+\; O(\varepsilon^6).
\end{align*}
A direct numerical comparison against the exact $M(E)$ relation confirms the expected $O(\varepsilon^6)$ convergence: the maximum residual falls from $2.1\times10^{-2}$ (first-order approximation alone) to $4.3\times10^{-3}$ (fifth order) at $e=2$, and from $5.0\times10^{-4}$ to $1.9\times10^{-5}$ at $e=5$ - a gain of a factor of about~25--27, consistent with a residual of order $\varepsilon^6$ rather than $\varepsilon^4$.

\section{Reconstructing Jacques Cassini's 1740 example}
\label{sec:appendix-jc}

Cassini's data are: greatest distance $AF=1\,016\,850$, least $FP=983\,150$, mean $1\,000\,000$, so $\varepsilon=(AF-FP)/2\,000\,000=0.01685$ and $k^2=AF\cdot FP$. For $FH=1\,010\,000$ the defining property $FH\cdot HE=k^2$ gives $HE=k^2/FH=989\,818$ (computed with more digits, $989\,817.8985$), exactly his value. The law of cosines in triangle $EHF$ gives $\widehat{EFH}=52^\circ26'28''$ and $\widehat{FEH}=126^\circ0'45''$; in the pericentre convention $\nu_{\mathrm{ov}}=127^\circ33'32''$, $M_{\mathrm{eq}}=126^\circ0'45''$, so $\nu_{\mathrm{ov}}-M_{\mathrm{eq}}=1^\circ32'47''$, his value. Solving Kepler's equation for the same $M$ and $\varepsilon$ gives $\nu-M=1^\circ32'33''$, so the discrepancy is $14.49''$ (his $14''$).

Repeating the computation for $M\in[0^\circ,180^\circ]$ gives a maximum of $14.887''$ at $M\approx132.9^\circ$. A perturbative expansion in $\varepsilon$ gives $\nu_{\mathrm{ov}}-M_{\mathrm{eq}}=2\varepsilon\sin M+\varepsilon^2\sin2M+O(\varepsilon^3)$ against the Keplerian $2\varepsilon\sin M+\tfrac54\varepsilon^2\sin2M+O(\varepsilon^3)$, so that the discrepancy is
\begin{equation}
(\nu_{\mathrm{ov}}-M_{\mathrm{eq}})-(\nu-M)=-\frac{\varepsilon^2}{4}\sin2M+O(\varepsilon^3),
\end{equation}
of amplitude $\varepsilon^2/4=14.64''$. The $O(\varepsilon^3)$ remainder is not negligible at this precision: the exact curve departs from the first term by up to $1.645''$ (and by $1.644''$ at $M=90^\circ$, where the first term vanishes), so the exact discrepancy is only approximately antisymmetric about $90^\circ$. The numbers above were recomputed independently, by a logarithm-free calculation, and agree with Cassini's printed values to the arcsecond.

\end{document}